\documentclass[10pt,a4paper]{article}
\usepackage{jheppub_kim}
\usepackage{pdflscape}
\usepackage{amsmath}
\usepackage{amssymb}
\usepackage{dcolumn}
\usepackage{bm}
\usepackage{color}
\usepackage{epsfig}
\usepackage{amsfonts}
\usepackage{graphicx}
\usepackage{subfigure}
\usepackage{dcolumn}

\begin{document}

\title{Constraints on Energy Momentum Squared Gravity from cosmic chronometers and Supernovae Type Ia data}

\author[a]{Chayan Ranjit}
\author[b]{Prabir Rudra,}
\author[c]{Sujata Kundu}

\affiliation[a]{Department of Mathematics, Egra S.S.B. College,
Egra, Purba Medinipur-721429, W.B., India}

\affiliation[b] {Department of Mathematics, Asutosh College,
Kolkata-700 026, India.}

\affiliation[c]{Department of IT, Narula Institute of Technology,
Kolkata-700109, W.B., India.}

\emailAdd{chayanranjit@gmail.com, chayan@associates.iucaa.in}
\emailAdd{prudra.math@gmail.com, rudra@associates.iucaa.in}
\emailAdd{sujatakundu10@gmail.com}

\abstract{In this work we perform an observational data analysis
on the energy momentum squared gravity model. Possible solutions
for matter density are obtained from the model and their
cosmological implications are studied. Some recent observational
data is used to constrain model parameters using statistical
techniques. We have used the cosmic chronometer and SNe Type-Ia
Riess (292) $H(z)-z$ data-sets in our study. Along with the
data-sets we have also used baryon acoustic oscillation (BAO) peak
parameter and cosmic microwave background (CMB) peak parameter to
obtain bounds on the model parameters. Joint analysis of the data
with the above mentioned parameters have been performed to obtain
better results. For the statistical analysis we have used the
minimization technique of the $\chi^{2}$ statistic. Using this
tool we have constrained the free parameters of the model.
Confidence contours have been generated for the predicted values
of the free parameters at the $66\%$, $90\%$ and $99\%$ confidence
levels. Finally we have compared our analysis with the union2 data
sample presented by Amanullah et al.,2010 and the recently
published Pantheon data sample. Finally a multi-component model is
investigated by adding dust to a general cosmological fluid with
equation of state $w=-1/3$. The density parameters were studied
and their values were found to comply with the observational
results.}


\maketitle

\section{Introduction}
Incorporating the late cosmic acceleration (Riess et al. 1998;
Perlmutter et al. 1999; Spergel et al. 2003) has been the greatest
challenge of theoretical cosmology. Since its discovery at the
turn of the last century it has continued to puzzle the greatest
minds of present time. It is known that General Relativity (GR) is
the most promising theory of gravity available to us given its
gradual success in complying with the observations over the years.
These include the perihelion precession of Mercury, gravitational
lensing, gravitational waves, etc. But in spite of all these GR
continues to remain inconsistent at cosmological distances because
it is unable to incorporate the late cosmic acceleration in its
framework. So the most logical solution to this will be to look
for suitable modifications of GR which will have provisions to
include the cosmic acceleration.

The possible modifications that have been attempted till now can
be categorized into two types. One category modifies the matter
content of the universe thus giving it an exotic nature which can
successfully drive the acceleration. This exotic matter which
generates an anti-gravitating stress is termed as \textit{dark
energy} (DE) basically due to its invisible nature. The first
attempt towards this modification was done by Einstein himself in
the form of cosmological constant ($\Lambda$). But this idea is
plagued by a some shortcomings, such as the cosmological constant
problem (Martin 2012). This problem deals with the mismatch
between the observed and the theoretical values of the
cosmological constant. Moreover there is also a cosmic coincidence
problem between dark energy and dark matter (Zlatev, Wang \&
Steinhardt 1999). An extensive review on DE was given by Brax
(2018).

The second category tries to modify the underlying spacetime
geometry without bothering about the matter content. In the second
category one introduces modifications to the spacetime curvature
proposed in the Einstein-Hilbert action such that the resulting
scenario admits the cosmic acceleration. This leads to the theory
of \textit{modified gravity}. Here the modification to Einstein
gravity is brought about at large distances specifically beyond
our solar system (Carroll et al. 2004, Cognola et al. 2008,
Clifton et al. 2012). Extensive reviews on this topic can be found
in literature (Nojiri, Odintsov \& Oikonomou 2017; Nojiri \&
Odintsov 2007). Now there can be various ways of modifying
Einstein's theory of GR. The most straightforward way seems to be
via the modification of the gravitational lagrangian (which is of
the form $\mathcal{L}_{\rm GR}= R$, $R$ being the scalar
curvature) of the Einstein-Hilbert action. The idea is to replace
this lagrangian by an arbitrary function of $R$. This gives rise
to the popular $f(R)$ theory of gravity, where the gravity
lagrangian is given by $\mathcal{L}_{f(R)}=f(R)$. It is obvious
that by choosing suitable functions $f(R)$ we can investigate the
non-linear effects of the scalar curvature in the evolution of the
universe. The degrees of freedom derived from this extended
functionality helps us to overcome the restrictions of GR.
Detailed reviews on $f(R)$ gravity was given by De Felice \&
Tsujikawa (2010) and Sotiriou \& Faraoni (2010).

Mathematically speaking any arbitrary functions of $R$ can be
considered in the gravitational lagrangian of the Einstein-Hilbert
action and its dynamics may be studied. But all such models may
not be cosmologically viable in the sense of their compliance or
non-compliance with observational datasets. In this connection
Amendola, Polarski \& Tsujikawa (2007) studied the cosmological
viability of $f(R)$ theories and ruled out some particular classes
of power law models. Coupling between matter and curvature is
another theoretical tool to modify gravity such as to generate
additional degrees of freedom which accounts for the accelerated
cosmic expansion. In this context a particular type of coupling
known as non-minimal coupling (NMC) (Azizi \& Yaraie 2014) have
produced very interesting results. Amendola et al. (2007) studied
the cosmological dynamics of $f(R)$ theories using a dynamical
system mechanism. A reconstruction scheme for $f(R)$ theories was
studied by Nojiri \& Odintsov (2006) and the dynamics of large
scale structure was investigated by Song, Hu \& Sawicki (2007).
Some generic formulations using $f(R)$ gravity was performed by
Capozziello, Laurentis \& Faraoni (2010).

A particular class of models where the gravitational Lagrangian is
formed out a generic function of the curvature scalar $R$ and the
trace of the stress-energy tensor $T$ has gained popularity in
recent times. In literature such models have been named as
$f(R,T)$ theories (Harko et al. 2011; Nagpal et al. 2019). Further
modifications to $f(R,T)$ theories are introduced by using a
scalar field which give rise to $f(R,T^{\phi})$ theories (Harko et
al. 2011), where $T^{\phi}$ is the trace of the stress energy of
the scalar field. Haghani et al. (2013) proposed a different type
of coupling between geometry and matter known by the name of
$f(R,T,R_{\mu\nu}T^{\mu\nu})$ gravity theory. This is a more
generic theory involving matter and geometry non-minimally coupled
to each other, where the gravitational Lagrangian depends of the
curvature scalar, the trace of the matter energy-momentum tensor,
and the contraction between the curvature tensor and the matter
energy-momentum tensor. Bertolami et al. (2007), Harko (2008),
Harko (2010) and Harko \& Lobo (2010), proposed some models
involving the matter Lagrangian density $L_m$ non-minimally
coupled to the curvature scalar $R$ giving rise to generic
$f(R,L_m)$ models. Under this gravity matter particles experience
extra force appearing in the direction orthogonal to the
four-velocity.

In continuation to the generalization of the above mentioned
$f(R,L_m)$ theories, we can further impose modifications by
including some analytic function of $T_{\mu\nu}T^{\mu\nu}$ (where
$T_{\mu\nu}$ is the stress energy-momentum tensor of the matter
component) in the gravity Lagrangian. Such modifications result in
$f(R,T_{\mu\nu}T^{\mu\nu})$ theories of gravity known as the
Energy-Momentum-Squared gravity (EMSG). Katirci \& Kavuk (2014)
proposed this theory as a covariant generalization of GR which
allows the existence of a term proportional to
$T_{\mu\nu}T^{\mu\nu}$ in the gravity Lagrangian. Cosmological
models in EMSG was studied by Board \& Barrow (2017). Using the
specific functional $f(R,T^2)=R+\eta T^2$, where $\eta$ is a
constant, Roshan \& Shojai (2016) investigated the possibility of
cosmological bounce. Bahamonde, Marciu \& Rudra (2019) explored
the cosmological dynamical system in the background of EMSG.
Energy-Momentum-Powered gravity (EMPG) (Board \& Barrow 2017,
Akarsu, Katirci \& Kumar (2018)) is a generalization of EMSG,
where we consider $f(R,T^2)=R+\eta (T^2)^n$, with $\eta$ and $n$
constant parameters. EMSG has proven to be a promising
cosmological theory and there has been a lot of interest in
studying its features. Matter wormholes of non-exotic nature are
explored in the background of EMSG by Moraes \& Sahoo (2018), and
possible parameter constraints from the observation of neutron
stars are discussed by Akarsu, Barrow, Cikintoglu, Eksi \& Katirci
(2018). Various cosmological features of EMSG are discussed by
Nari \& Roshan (2018), Akarsu, Katirci, Kumar, Nunes \& Sami
(2018) and Keskin (2018). In the background of
Energy-Momentum-Powered (EMPG) theories (a generalization of EMSG
theory), the late time cosmic acceleration have been investigated
by Akarsu, Katirci \& Kumar (2018), considering the a
pressure-less fluid. In connection to EMSG theories one more
theory by the name of energy-momentum-Log gravity (EMLG) was
introduced by Akarsu et al. (2019). Here a specific form of
logarithmic function was introduced as given by,
$f(T_{\mu\nu}T^{\mu\nu})=\alpha \ln\left(\lambda
T_{\mu\nu}T^{\mu\nu}\right)$, where $\alpha$ and $\lambda$ are
constants.

The most obvious question that arises in any form of modification
of gravity is that what should be the choice for the arbitrary
functions present in the model. This requirement is equivalent to
constraining the model parameters, thus constraining the model as
a whole. The answer to this comes from various sources. The
initial constraining comes from heuristic and theoretical
arguments such as the requirement of a ghost free theory that
possesses stable perturbations (De Felice \& Tsujikawa 2010).
Requiring that the theory possesses Noether symmetries is another
way to constrain a model (Paliathanasis, Tsamparlis \& Basilakos
2011; Paliathanasis 2016). However in order to further constrain a
model we need to use observational data, which can constrain a
model to any degree of accuracy via statistical procedures and
algorithms. For the EMSG theory this study is almost absent in
literature. Akarsu, Katirci \& Kumar (2018) constrained the
parameter space of the corresponding EMPG model (a model related
to the EMSG model) by relying on various values of the Hubble
parameter $(H(z))$ (where $z$ is the redshift) reported by Farooq
\& Ratra (2013). In this paper we would like to constrain the
corresponding model parameters of EMSG theory using $z-H(z)$
observational data taking into consideration the BAO and CMB peak
parameters. We would like to use the well-known cosmic chronometer
data-set (Moresco et al. 2016; Gomez-Valent \& Amendola 2019) and
the Supernova Type Ia 292 data set (Riess et al. 2004, 2007;
Astier et al. 2006) for our analysis.

The paper is organized as follows: Section II deals with the basic
equations of energy momentum squared cosmology. In section III we
present a detailed observational data analysis mechanism using a
particular model. Section IV is dedicated to the study of a
multi-component universe model. Finally the paper ends with a
discussion and some concluding remarks in section V.

\section{Energy-momentum Squared cosmology}

The action of the model is given by (Katirci \& Kavuk 2014)
\begin{equation}
S=\frac{1}{2\kappa^2}\int d^4x \sqrt{-g} f(R,\mathbf{T^2})  +
S_{\rm m},
\end{equation}
where $f(R,\mathbf{T^2})$ is an analytic function of the square of
the energy-momentum tensor $\mathbf{T^2}=T^{\mu\nu}T_{\mu\nu}$ and
the Ricci scalar $R$. Here, $\kappa^2=8\pi G$ and $S_{\rm m}$ is
the matter action.

By varying the action with respect to the metric we get the
following field equations
\begin{equation}
R_{\mu \nu}f_R  +g_{\mu\nu} \Box f_R-\nabla_{\mu}\nabla_{\nu}
f_R-\frac{1}{2} g_{\mu\nu}f=\kappa^2 T_{\mu \nu}-f_{\mathbf{T^2}}
\Theta_{\mu \nu}\,, \label{FieldEq}
\end{equation}
where $\Box=\nabla_\mu \nabla^\mu$, $f_{R}=\partial f/\partial R$,
$f_{\mathbf{T^2}}=\partial f/\partial \mathbf{T^2}$ and
\begin{equation}
    \Theta_{\mu\nu}=\frac{\delta (\mathbf{T^2})}{\delta g^{\mu\nu}}= \frac{\delta (T^{\alpha\beta}T_{\alpha\beta})}{\delta g^{\mu\nu}}=-2L_{\rm m}\Big(T_{\mu\nu}-\frac{1}{2}g_{\mu\nu}T\Big)-T\, T_{\mu\nu}+2T^{\alpha}_{\mu}T_{\nu\alpha}-4T^{\alpha\beta}\frac{\partial^2 L_{\rm m}}{\partial g^{\mu\nu}\partial g^{\alpha\beta}}\,,\label{Theta}
\end{equation}
where $T$ is the trace of the stress-energy tensor. Taking
covariant derivatives of the field equation~\eqref{FieldEq}, we
get the following conservation equation
\begin{eqnarray}
\kappa^2\nabla^\mu T_{\mu\nu}=-\frac{1}{2}g_{\mu\nu}\nabla^\mu
f+\nabla^\mu(f_{\mathbf{T}^2}\Theta_{\mu\nu})\,.\label{conservation1}
\end{eqnarray}
It is clear from the above equation that, the standard
conservation equation does not hold for this theory. If one
chooses $f(R,\mathbf{T}^2)=2\alpha \log(\mathbf{T}^2)$, one
obtains the same result given in Akarsu et al. (2019).

Now we will consider the flat FLRW cosmology for this model whose
metric is described by
\begin{equation}
ds^2=-dt^2+a^2(t)\delta_{ik}dx^idx^k,
\end{equation}
with $\delta_{ik}$ is the Kronecker symbol and $a(t)$ the scale
factor. We consider that the matter content is given by a standard
perfect fluid with $T_{\mu\nu}=(\rho+p)u_\mu u_\nu + p g_{\mu\nu}$
with $u_{\mu}$ being the 4-velocity and $\rho$ and $p$ are the
energy density and the pressure of the fluid respectively. Using
these, the energy-momentum tensor gives us $T^2=\rho^2+3p^2$.
Further, let us consider $L_{\rm m}=p$ which permits us to rewrite
$\Theta_{\mu\nu}$ defined in eqn. \eqref{Theta} as a quantity
which does not depend on the function $f$, as given below (Board
\& Barrow 2017)
\begin{equation}
\Theta_{\mu\nu}=-\Big(\rho^2+4 p\rho+3p^2\Big)u_\mu u_\nu\,.
\end{equation}
The modified FLRW equations compatible with this particular action
are given by
\begin{eqnarray}
-3f_R\Big(\dot{H}+ H^2\Big)+\frac{f}{2}+3 H \dot{f_R}&=&\kappa^2\Big(\rho+\frac{1}{\kappa^2}f_{\mathbf{T^2}}\Theta^2\Big)\,,\label{FW1} \\
-f_R(\dot{H}+3 H^2)+\frac{1}{2}f+\ddot{f_R}+2 H
\dot{f_R}&=&-\kappa^2 p\,,\label{FW2}
\end{eqnarray}
where 'dot' denote differentiation with respect to time $t$ and
$H=\dot{a}/a$ is the Hubble parameter, and
\begin{equation}
\mathbf{\Theta^2}:=\Theta_{\mu\nu}\Theta^{\mu\nu}=\rho^2+4p
\rho+3p^2\label{Theta2}
\end{equation}
was defined. The conservation equation \eqref{conservation1} for
this model is given by,
\begin{eqnarray}
\kappa^2(\dot{\rho}+3H(\rho+p))&=&-\mathbf{\Theta^2}\dot{f}_{\mathbf{T^2}}-f_{\mathbf{T^2}}
\Big[3 H\mathbf{\Theta^2}+\frac{d}{dt}\Big(2\rho
p+\frac{1}{2}\mathbf{\Theta^2}\Big)\Big]\,.\label{conservation3}
\end{eqnarray}
It is quite clear that the standard conservation equation is
violated for EMSG cosmology for an arbitrary function
$f(R,\mathbf{T^2})$. If one chooses $f(R,\mathbf{T^2})=f(R)$, all
the terms on the RHS of the above equation are zero and the
standard conservation equation is recovered.

We can rewrite the modified FLRW equations as,
\begin{eqnarray}
    3H^2&=&\kappa^{2}\rho_{eff}=\kappa^2(\rho+\rho_{\rm modified})\,,\\
      3H^2+2\dot{H}&=&-\kappa^{2}p_{eff}=-\kappa^2(p+p_{\rm modified})\,,
\end{eqnarray}
where we can define the energy density and pressure for the EMSG
modifications as
\begin{eqnarray}
\rho_{\rm modified}&=&-\frac{1}{f_R}\left[\rho+\frac{1}{\kappa^{2}}\left\{f_{T^2}\left(\rho^{2}+4p\rho+3p^{2}\right)-\frac{f}{2}-3H\dot{f_R}+3\dot{H}f_{R}\right\}\right]-\rho\,,\label{rho}\\
p_{\rm
modified}&=&-\left[\frac{1}{f_R}\left\{p+\frac{1}{\kappa^{2}}\left(\frac{f}{2}+\ddot{f_{R}}+2H\dot{f_{R}}\right)\right\}+\frac{\dot{H}}{\kappa^{2}}\right]-p\,.\label{p}
\end{eqnarray}

For the matter fluid we will consider a standard barotropic
equation of state given by,
\begin{equation}\label{pressure1}
    p=w \rho\,,
\end{equation}
where $w$ is the equation of state (EoS) parameter. Using this
relation one gets that
\begin{equation}\label{theta1}
\mathbf{\Theta^2}=(1+4w+3w^2)\rho^2\,,
\end{equation}
and then the conservation equation \eqref{conservation3} becomes
\begin{eqnarray}
\dot{\rho}+3H(w+1)\rho&=&-f_{\mathbf{T^2}} \left[3 \left(3 w^2+4 w+1\right) H \rho ^2+\left(3 w^2+8 w+1\right) \rho \dot{ \rho}\right]\nonumber\\
&&-\left(3 w^2+4 w+1\right) \rho^2
\dot{f}_{\mathbf{T^2}}\,.\label{conservation4}
\end{eqnarray}
It should be noted that this theory does not satisfy the standard
continuity equation given below
\begin{equation}\label{origcont}
\dot{\rho}+3H\left(w+1\right)\rho=0
\end{equation}
The covariant divergence of the field equations produce non-zero
terms on the right hand side of the Eq.\eqref{origcont}, thus
leading to the modified continuity equation given by
Eq.\eqref{conservation4}. Finally we can define the effective
equation of state (EOS) as,
\begin{equation}
w_{eff}=\frac{p_{eff}}{\rho_{eff}}= \frac{w\rho+p_{\rm
modified}}{\rho+\rho_{\rm modified}}\,.\label{eos2}
\end{equation}

\subsection{Integrating the Continuity equation}
Integrating the continuity equation \eqref{conservation4} is not
at all straightforward and quite a difficult task for this model.
The obvious reason being the non-standard nature of the continuity
equation satisfied by this theory which has been already discussed
before. The non-zero term on the right hand side of the modified
continuity equation \eqref{conservation4} poses a real
mathematical challenge for this operation.  In this subsection we
would like to integrate the continuity equation independent of any
particular model and express the density parameter $\rho$ in terms
of the redshift parameter $z$. We see that the conservation
equation \eqref{conservation4} is not integrable for any value of
$w$ by the known mathematical methods. It is integrable for only
$w=-1/3$ and $w=-1$ (Board \& Barrow 2017). We know that $w=-1$
corresponds to the $\Lambda CDM$ cosmology and $w<-1/3$ indicates
quintessence (dark sector). On solving eqn.\eqref{conservation4}
for $w=-1$, we get two real solutions for the density parameter
$\rho$ given by,
\begin{equation}\label{cons1}
\rho=\frac{1}{f_{\mathbf{T^2}}}~~~~~~~~~~~   and
~~~~~~~~~~\rho=C_{0}
\end{equation}
where $C_{0}$ is a constant. For $w=-1/3$ we get only one real
value for $\rho$ given by,
\begin{equation}\label{cons2}
\rho=-\frac{3W\left[\frac{4}{3}\left\{-e^{-C_1}
(f_{\mathbf{T^2}})^3
\left(z+1\right)^6\right\}^{1/3}\right]}{4f_{\mathbf{T^2}}}
\end{equation}
where $W[y]$ is the Lambert $W$ function, $z$ is the redshift
parameter and $C_{1}$ is the integration constant. Lambert $W$
function returns the value $x$ that solves the equation
\begin{equation}\label{lambert}
y=x Exp(x)
\end{equation}

In the next section, an observational data analysis will be
performed on the gravity model. Specific models will be considered
for the analysis and the model parameters will be constrained
using observational data. Hereafter, we will use geometric units
$\kappa=1$. Using the expression for density we obtain the
deceleration parameter for the model $f(R,\mathbf{T^2})=R+\eta
\mathbf{T^2}$ as,
\begin{equation}
q=-\frac{W\left[\frac{4}{3} \left\{-e^{-C_1} \eta ^3
(z+1)^6\right\}^{1/3}\right]}{W\left[\frac{4}{3} \left\{-e^{-C_1}
\eta ^3 (z+1)^6\right\}^{1/3}\right]+2}
\end{equation}
By choosing a particular model of EMSG one can check the evolution
of $q$ with respect to $z$ using the constrained values of the
model parameters $\eta$ and $C_{1}$. Negative values of $q$ will
indicate the accelerated expansion of the universe and the
viability of the model will be established.

\subsection{Cosmological Implication of the solution}

In the Eqns.\eqref{cons1} and \eqref{cons2}, we have got the
solutions for the energy density by solving the continuity
equation \eqref{conservation4} for two different equation of
states. The solution obtained in Eq.\eqref{cons1} seems to be very
trivial in nature. With a model of the form
$f(R,\mathbf{T^2})=R+\eta \mathbf{T^2}$, we see that the solution
given by eqn.\eqref{cons1} become $\rho=1/\eta$ or $\rho=C_{0}$.
Here $\rho=C_{0}$ is a physically unrealistic solution because a
constant density parameter will imply a static universe. Similarly
if $\eta$ is a positive or negative constant, $\rho=1/\eta$ will
be a constant value and will not represent a realistic solution.
However since $\eta$ represents the gravitational coupling
strength of the modification to gravity, it will be no harm in
considering it as a time dependent variable $\eta(t)$. In such a
scenario, we see that with the increase of $\eta$ with time $t$,
there is a corresponding depletion in the value of $\rho$ which is
the ideal scenario for an expanding universe. We see that if
$\eta(t)$ is a monotonically increasing function of time, then the
density of universe gradually decreases and finally for late time
universe, as $t\rightarrow +\infty$, $\eta(t)\rightarrow +\infty$,
then $\rho \rightarrow 0$. When $t \rightarrow 0$ (corresponding
to big bang), we see that $\eta(t) \rightarrow 0$ for properly
adjusted parameterization, and consequently $\rho \rightarrow
+\infty$. Now coming to the solution given in eqn.\eqref{cons2},
corresponding to $w=-1/3$, we see that the expression for $\rho$
is given by a special function (Lambert W function). Now there are
countably many branches of the W function, which are denoted by
$W_{k}(z)$ for integral $k$. $W_{0}(z)$ is called the principal
branch, with $W_{0}(0)=0$. Now if we consider this principal
branch solution, we see that for the model considered in the
previous case, as $\eta \rightarrow 0$, we see that $\rho$ takes
an indeterminate form. So using the L'Hospital's rule and also the
result $W_{0}'(0)=1$, we get $\rho \rightarrow 3/4$ as $\eta
\rightarrow 0$. This value corresponds to GR for $t=1$ following
the work of Akarsu et al 2018. So we see that by proper scaling of
the solution or via proper fine tuning of the initial conditions
there is a chance of realizing the standard radiation dominated
universe of GR from the solution that we have got. Moreover since
the W function is quite complicated and not monotonic in nature,
we cannot comment on the cases $\eta>0$ and $\eta<0$. We consider
this as an immediate disadvantage of this solution. But we are
hopeful that by proper Taylor series approximation of the W
function we can realize proper cosmological scenario which
corresponds to observations. So overall we should say that our
obtained solution is quite promising as far as its cosmological
implication is concerned.

\section{Observational Data Analysis mechanism}
In this section we intend to perform an observational data
analysis to investigate the acceptable range of model parameters
using observational data. Using the modified FLRW equations
\eqref{FW1}, \eqref{FW2}, \eqref{pressure1} and \eqref{theta1} we
get
\begin{equation}\label{hobs}
H=\frac{\sqrt{24f_{R}\left(3\ddot{f_R}+f+f_{\mathbf{T^2}}
\left(\rho^2+3\rho^{2}w^{2}+4\rho^{2}w\right)+\rho+3\rho
w\right)+9\dot{f_R}^2}+3\dot{f_R}}{12f_{R}}
\end{equation}
Now we have to consider a particular EMSG model to proceed
further.


\subsection{The Model}
Here we consider a particular EMSG model given by Katirci \& Kavuk
(2014) and Roshan \& Shojai (2016),
\begin{equation}\label{emsg1}
f(R,\mathbf{T^2})=R+\eta \mathbf{T^2}
\end{equation}
where the Ricci scalar $R$ is given by
$R=6\left(\dot{H}+2H^{2}\right)$ and $\eta$ is a constant which
can be both positive or negative. This is the simplest model of
the EMSG gravity where both $R$ and $T^2$ are present. From the
work of Roshan \& Shojai (2016) we see that the values of $\eta$
are constrained to lie in the negative region to give satisfactory
cosmological behaviour. Using the expression of $\rho$ from
eqn.\eqref{cons2} for $w=-1/3$ in eqn.\eqref{hobs} we get the
following expression for the Hubble parameter for the considered
model,
\begin{equation}\label{hobs2}
H(z)=\frac{1}{2\sqrt{2}}\left[\sqrt{-\frac{W\left[\frac{4}{3}
\left\{-e^{-C_{1}}\eta^{3}(z+1)^6\right\}^{1/3}\right]}{\eta}}
\sqrt{W\left[\frac{4}{3}\left\{-e^{-C_{1}}\eta^{3}(z+1)^6\right\}^{1/3}\right]+2}\right]
\end{equation}
Here we have expressed the Hubble parameter $H(z)$ in terms of the
redshift parameter $z$. This will serve as our theoretical
framework in this study. The idea is to use $z-H(z)$ data-sets to
constrain the model parameters using statistical techniques. This
will give an idea about the viability of EMSG as a theory of
gravity and its applicability to cosmology when compared with a
standard cosmological model like the $\Lambda CDM$.




\subsection{The Data}
Here, we have used two $z-H(z)$ data sets namely the Cosmic
Chronometer (CC) (Jimenez \& Loeb 2002; Moresco 2015; Simon, Verde
\& Jimenez 2005; Stern et al. 2010; Zhang et al. 2014) \&
Supernovae Type Ia (SNe Type-Ia) Riess $292$ (Riess et al. 2004,
2007; Astier et al. 2006). The CC data (which is a $30$ point data
set) is given in Table (\ref{T1}) in Appendix Section. The SNe
Type-Ia is a $292$ points data set compiled from the references
(Riess et al. 2004, 2007; Astier et al. 2006). To keep the paper
compact we have not included it directly over here. The reader may
refer to the references to see the exact data points. From the
table it can be seen that along with the values of $z$ and $H(z)$
we also have the corresponding values of $\sigma(z)$ which
represents the standard deviation of the particular data point.
This gives an idea about the deviation of the point about the mean
or the best fit line, which arises from the error in measurement.

The cosmic chronometers are a very powerful set of tools in
understanding the evolution of the universe, which was first given
by Jimenez \& Loeb (2002). This data-set is actually a set of
values of Hubble parameter obtained at different redshifts
extracted through the differential age evolution of the passively
evolving early-type galaxies. We know that for FRW universe the
Hubble parameter can be expressed as
$H=-\left(1+z\right)^{-1}dz/dt$. From this it is obvious that by
measuring the gradient $dz/dt$ it is possible to measure the
Hubble parameter values and correspondingly retrieve the Hubble
data. A detailed description of the cosmic chronometer data can be
found in Moresco et al. (2016). As mentioned earlier the CC data
contains $30$ $z-H(z)$ data points spanning the redshift range
$0<z<2$. This data, obtained by the CC approach approximately
covers about $10 ~Gyr$ of cosmic time.

Supernovas are very bright events in sky and so they serve as
standard candles in astrophysical and cosmological observations.
Observations from Supernova Type Ia served as evidence for the
cosmic acceleration at the turn of the last century. To date
numerous supernova type Ia observations have been recorded in the
literature by various research collaborations. Here we are working
with the 292 points $z-H(z)$ data reported by Reiss et al. (Riess
et al. 2004, 2007; Astier et al. 2006).

\subsection{Analysis with Cosmic Chronometer \& Supernovae Type Ia Riess 292 ($H(z)$-$z$) Data Set}
\begin{figure}
\begin{minipage}[t]{0.5\linewidth}
\centering
\includegraphics[width=0.75\textwidth]{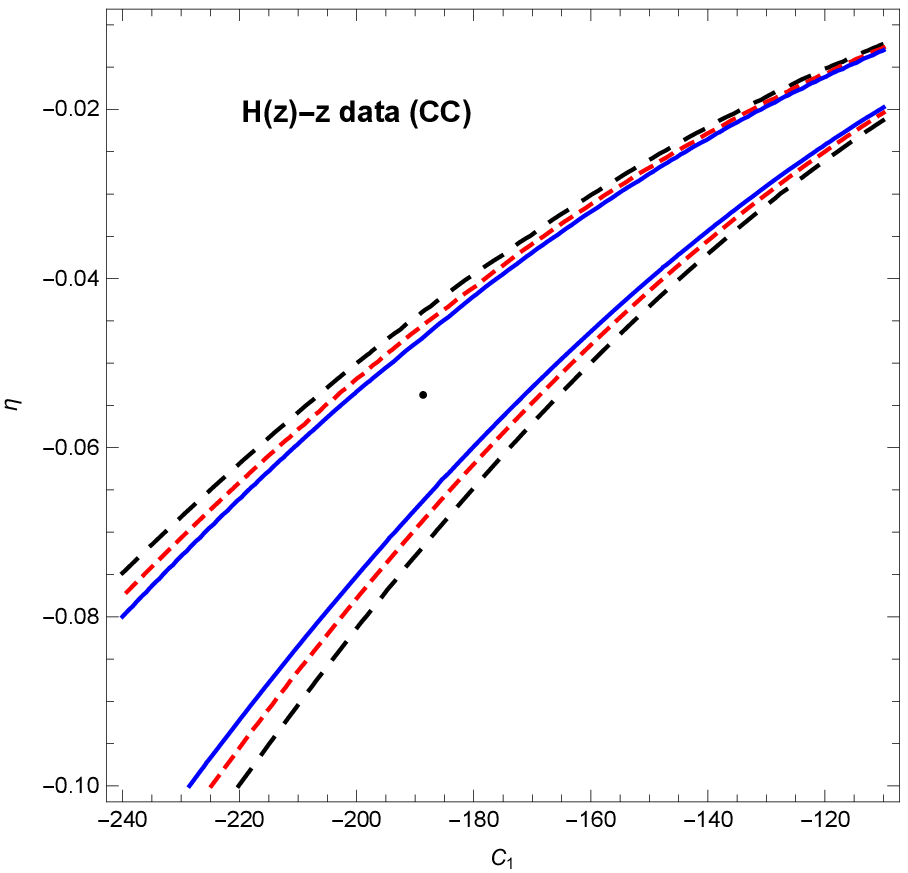}
\caption{Confidence contours for $\eta$ vs $C_1$ for the CC data
set.}\label{f1.1} \vspace{2mm}
\end{minipage}~~
\begin{minipage}[t]{0.5\linewidth}
\centering
\includegraphics[width=0.75\textwidth]{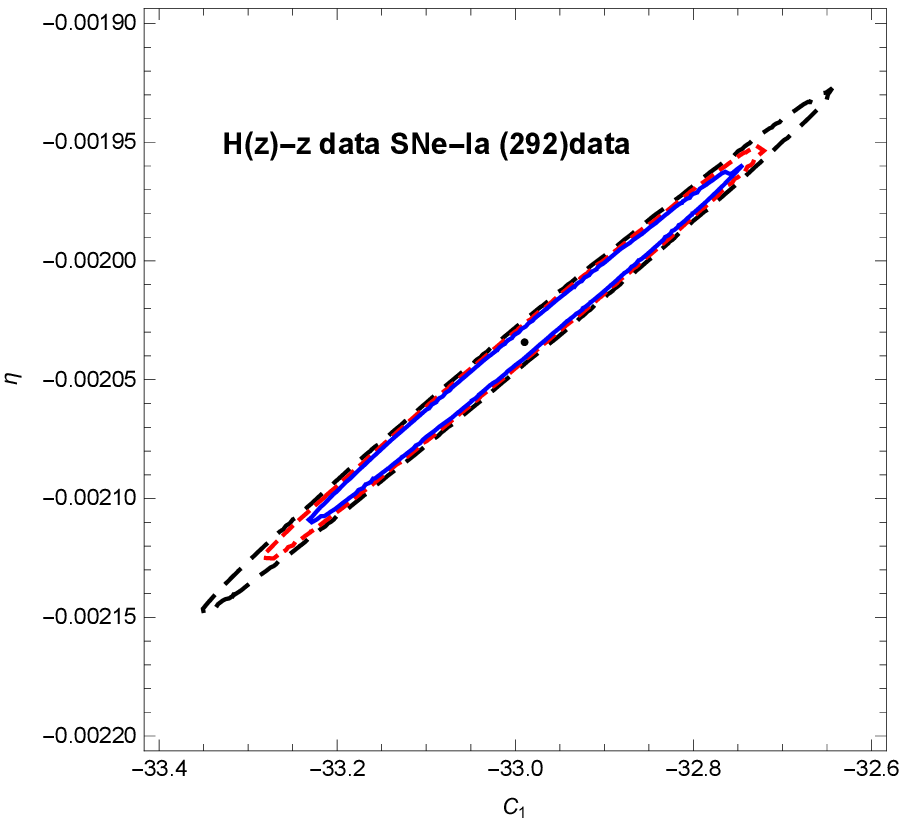}
\caption{Confidence contours for $\eta$ vs $C_1$ for the 292 SNe
Type-Ia data set.}\label{f1.2} \vspace{4mm}
\end{minipage}~~\\
\vspace{2mm}

Figure \ref{f1.1} shows the contourplot of $\eta$ against $C_1$
for the cosmic chronometer data set and Figure \ref{f1.2} shows
the contours of $\eta$ vs $C_1$ for the 292 SNe Type-Ia data set.
\end{figure}

Using the values of Hubble parameter from observational data at
different red-shifts given by the CC and Supernovae Type Ia data
sets, we want to analyze our predicted model. The best fitted
cosmological scenario with statistical errors is achieved through
an iterative chi-square ($\chi^{2}$) minimization technique. For
this purpose we will first establish the $\chi^{2}$ statistic as a
sum of standard normal distribution as follows:
\begin{equation}\label{Eqn1.5}
{\chi}_{CC/SNe-Ia}^{2}=\sum\frac{(H(z)-H_{obs}(z))^{2}}{\sigma^{2}(z)}
\end{equation}
where $H(z)$ and $H_{obs}(z)$ are the theoretical and
observational values of Hubble parameter at different red-shifts
respectively and $\sigma(z)$ is the corresponding error in
measurement of the data point. Here, $H_{obs}$ is a nuisance
parameter and can be easily marginalized. The present value of
Hubble parameter is considered as $H_{0}$ = 72 $\pm$ 8 Km s$^{-1}$
Mpc$^{-1}$ and we also consider a fixed prior distribution for it.
In this work, we intend to determine the range of model parameters
$\eta$ and $C_{1}$ of the EMSG model by minimizing the above
mentioned ${\chi}_{CC/SNe-Ia}^{2}$ statistic. The reduced chi
square can be written as
\begin{equation}\label{Eqn1.5.172}
L=\chi_{R}^{2}= \int
e^{-\frac{1}{2}{\chi}_{CC/SNe-Ia}^{2}}P(H_{0})dH_{0}
\end{equation}
where $P(H_{0})$ is the prior distribution function for $H_{0}$.
In order to represent the acceptable ranges of the model
parameters, we have generated the contour plots for 66\% (solid,
blue), 90\% (dashed, red) and 99\% (dashed, black) confidence
levels which are shown in figures \ref{f1.1} and \ref{f1.2} for
the two different data sets. From the figures we see that only
negative values of the parameters are allowed for this constrained
model with the given data. This observation is consistent with the
work of Roshan \& Shojai (2016). There the authors have shown that
only negative values of $\eta$ are cosmologically meaningful.
Negative values of $\eta$ leads to a bounce at early times and to
a satisfactory cosmological behaviour after the bounce. Basically,
this analysis provides observational data support to the
theoretical model and also predicts the acceptable ranges of the
parameters $\eta$ and $C_{1}$. The black dot in the contours
represent the best possible values of the parameters, as predicted
from our set-up, which is displayed in the table \ref{T3}.

\begin{table}[h!]
\centering \caption{CC $\&$ SNe Type-Ia data}\label{T3}
\begin{tabular}{|c|c|c|c|}
\hline
~~~~~~~$Data Type$~~~~~& ~~~~~~$\eta$ ~~~~~& ~~~~~~~$C_{1}$ ~~~~~~~~& ~~~$\chi^{2}_{min}$~~~~~~\\
    \hline
    CC data & -0.0537602 & -188.644 & 117.017 \\
     \hline
    SNe Type-Ia data &-0.0020342 & -32.9896 & 7963.69 \\
    \hline
\end{tabular}

\vspace{4mm} \textbf{Table\ref{T3}:} Shows the best fit values of
$\eta$, $C_1$ and the minimum values of $\chi^{2}$ for CC Data and
SNe Type-Ia data.
\end{table}

\subsection{Joint Analysis with CC data $+$ BAO \& Supernovae Type Ia Riess 292 data + BAO}
\begin{figure}
\begin{minipage}[t]{0.5\linewidth}
\centering
\includegraphics[width=0.75\textwidth]{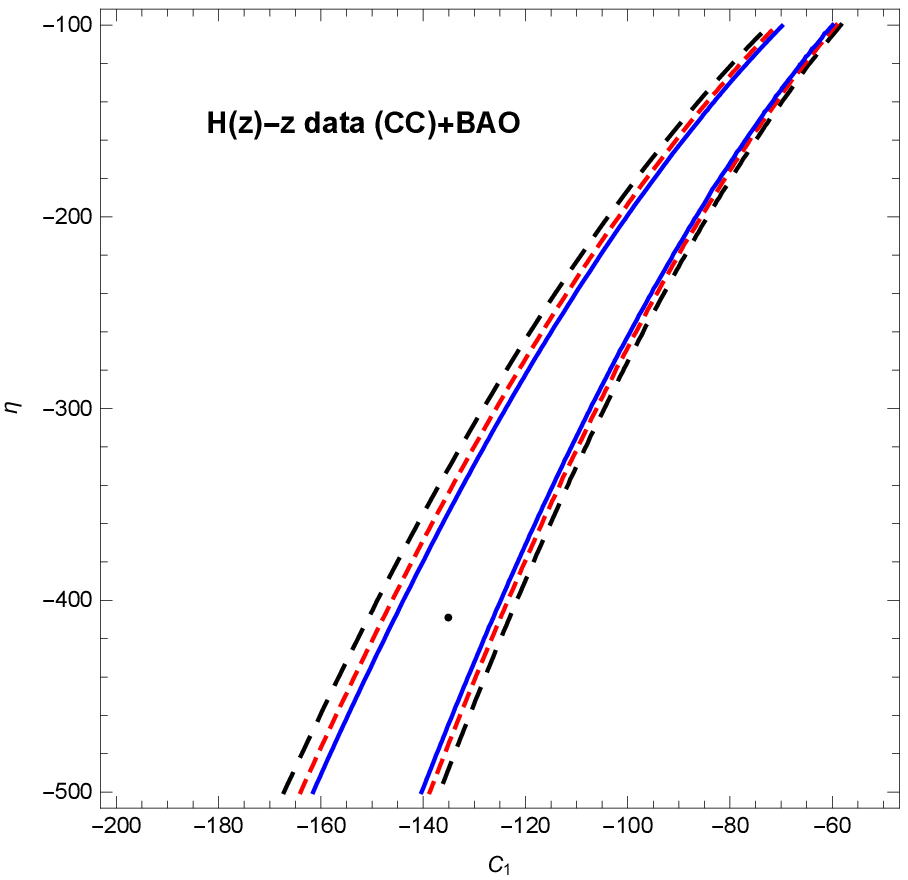}
\caption{Confidence contours for $\eta$ vs $C_1$ for the CC + BAO
data set.}\label{f2.1} \vspace{2mm}
\end{minipage}~~
\begin{minipage}[t]{0.5\linewidth}
\centering
\includegraphics[width=0.75\textwidth]{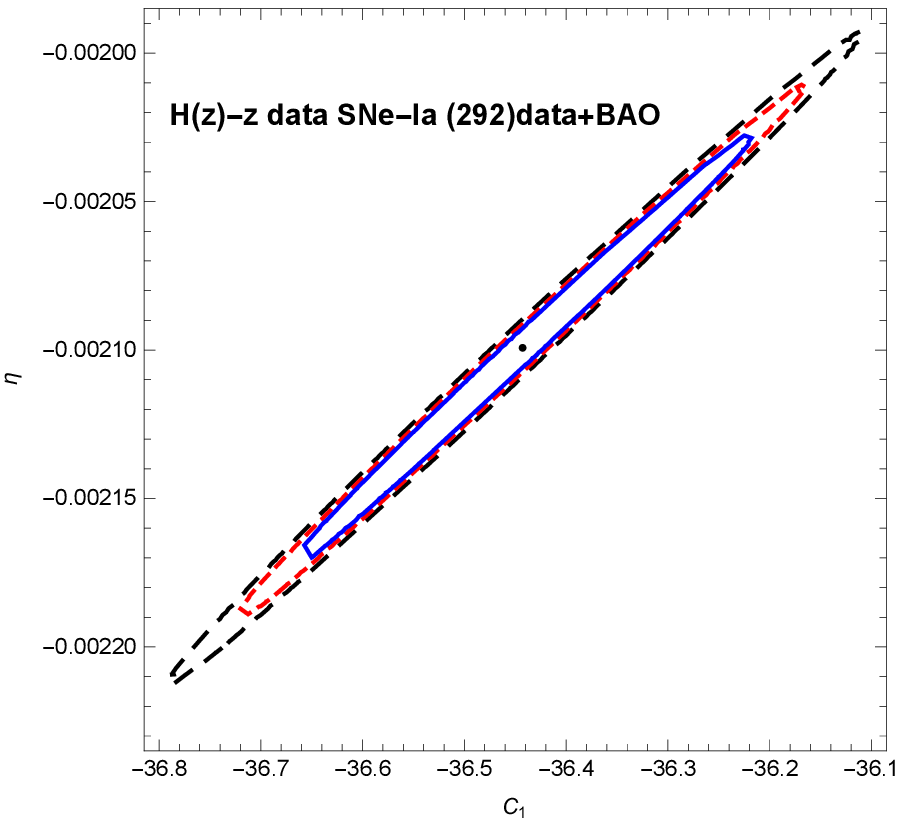}
\caption{Confidence contours for $\eta$ vs $C_1$ for the 292 SNe
Type-Ia + BAO data set.}\label{f2.2} \vspace{4mm}
\end{minipage}~~\\
\vspace{2mm}

Figure \ref{f2.1} shows the confidence contours of $\eta$ against
$C_1$ for the CC + BAO data set and Figure \ref{f2.2} shows the
corresponding confidence intervals of $\eta$ vs $C_1$ for the 292
SNe Type-Ia + BAO data set.
\end{figure}

Eisenstein et al. (2005) proposed a method of joint analysis of
observational data with the Baryon Acoustic Oscillation (BAO) peak
parameter in order to constrain cosmological toy models. The Sloan
Digital Sky Survey (SDSS) was one of the first red-shift survey by
which the BAO signal was directly detected at a scale of around
100 MPc. In this survey, spectroscopic samples from around 46748
luminous red galaxies were collected which spanned over 3816
square-degrees of sky and with an approximately diameter of five
billion light years. One of the major utility of this survey was
that, these findings confirmed the results obtained from Wilkinson
Microwave Anisotropy Probe (WMAP) using the sound horizon in the
present universe. Exploring the SDSS catalog we can retrieve a
such a distribution of matter existing in the universe, using
which we can easily probe a BAO signal by investigating the
presence of a relatively greater number of galaxies separated at
the sound horizon. The methodology employed in this above
mentioned analysis basically uses a blend of the angular diameter
distance parameter and the Hubble parameter at that red-shift
range. It is noteworthy that this analysis is self-reliant and
does not depend on the measurement of the present value of Hubble
parameter $H_{0}$. Moreover it does not contain any particular
dark energy model and in this sense it is fairly generic in
nature. Here we have examined the parameters $\eta$ and $C_{1}$ of
our model from the measurements of the BAO peak for low red-shift
(with range $0<z<0.35$) using standard $\chi^{2}$ analysis. The
error in the measurement is given by the standard deviation in the
data table, which can be fairly modelled by Gaussian distribution.
It is found that the low-redshift distance measurements are quite
reliant on the values of various cosmological parameters of the
model like the equation of state of dark energy. Moreover such
measurements do have the ability to quantify the current value of
Hubble parameter $H_{0}$ directly. The BAO peak parameter may be
defined by (Thakur, Ghose \& Paul 2009; Paul, Thakur \& Ghose
2010; Paul, Ghose \& Thakur 2011; Ghose, Thakur \& Paul 2012):
\begin{equation}\label{Eqn1.5.173}
{\cal A}=\frac{\sqrt{\Omega_{m}}}{E(z_{1})^{1/3}}
\left(\frac{1}{z_{1}}~\int_{0}^{z_{1}}
\frac{dz}{E(z)}\right)^{2/3}
\end{equation}
In the above expression $E(z)=H(z)/H_{0}$ is called the normalized
Hubble parameter. It is known from the SDSS survey that the
red-shift  $z_{1}=0.35$ is the prototypical value of red-shift
which we will consider in our analysis. The integral appearing in
the above expression is the dimensionless co-moving distance at
the red-shift $z_{1}$. From the observational data we have
motivated values for the above BAO peak ${\cal A}$. For the flat
model of the universe its value is estimated to be ${\cal
A}=0.469\pm 0.017$ using SDSS data (Eisenstein et al. 2005)
retrieved by probing a large group of luminous red galaxies. Now
the $\chi^{2}$ function for the BAO measurement can be written as

\begin{equation}\label{Eqn1.5.174}
\chi^{2}_{BAO}=\frac{({\cal A}-0.469)^{2}}{(0.017)^{2}}
\end{equation}
The total joint data analysis CC+BAO \& SNe Type-Ia+BAO for the
$\chi^{2}$ function may be defined by (Thakur, Ghose \& Paul 2009;
Paul, Thakur \& Ghose 2010; Paul, Ghose \& Thakur 2011; Ghose,
Thakur \& Paul 2012; Wu \& Yu 2007)
\begin{equation}\label{Eqn1.5.175}
\chi^{2}_{total}=\chi^{2}_{CC/SNe-Ia}+\chi^{2}_{BAO}
\end{equation}
The best fit values of the model parameters $\eta$ and $C_{1}$
according to our analysis via the joint scheme of CC data and BAO
peak are presented in Table \ref{T4}. Finally we have drawn the
contours for the predicted values of the parameters for the 66\%
(solid, blue), 90\% (dashed, red) and 99\% (dashed, black)
confidence intervals and depicted them in the figures \ref{f2.1}
and \ref{f2.2}. Just like the previous analysis here also we see
that only negative values of the parameters are allowed to realize
a cosmologically feasible scenario.

\begin{table}[h!]
\centering \caption{CC + BAO \& SNe Type-Ia + BAO Data}\label{T4}
\begin{tabular}{|c|c|c|c|}
\hline
~~~~$Data Type$ ~~~~~&~~~~$\eta$ ~~~~~& ~~~~~~~$C_{1}$ ~~~~~~~~& ~~~$\chi^{2}_{total-min}$~~~~~~\\
    \hline
   CC + BAO Data  & -408.965 & -135.083 & 1874.33 \\
    \hline
    SNe Type-Ia + BAO Data & -0.00209925 & -36.4428 & 8684.41 \\
    \hline
    \end{tabular}

\vspace{4mm} \textbf{Table\ref{T4}:}  Shows the best fit values of
$\eta$, $C_1$ and the minimum values of the $\chi^{2}$ statistic
for CC + BAO and SNe Type-Ia + BAO data.
\end{table}

\subsection{Joint Analysis with CC $+$ BAO $+$ CMB Data Sets \& Supernovae Type Ia Riess 292 Data + BAO + CMB}
\begin{figure}
\begin{minipage}[t]{0.5\linewidth}
\centering
\includegraphics[width=0.75\textwidth]{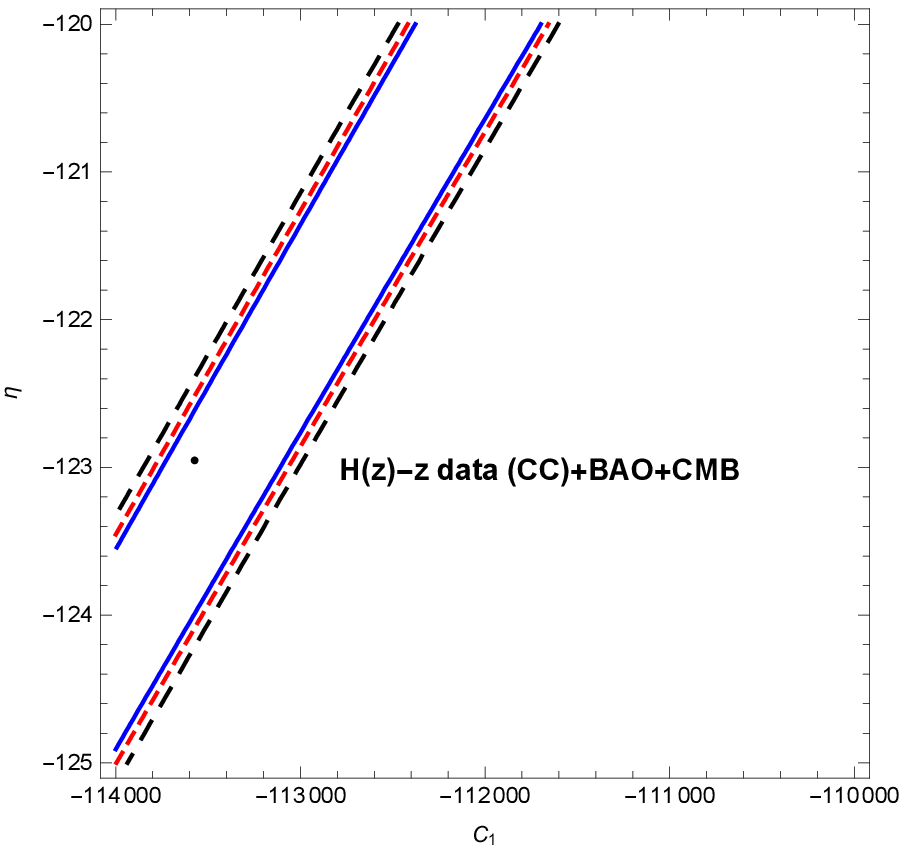}
\caption{Confidence contours for $\eta$ vs $C_1$ for the CC +BAO
+CMB data. }\label{f3.1} \vspace{2mm}
\end{minipage}~~
\begin{minipage}[t]{0.5\linewidth}
\centering
\includegraphics[width=0.75\textwidth]{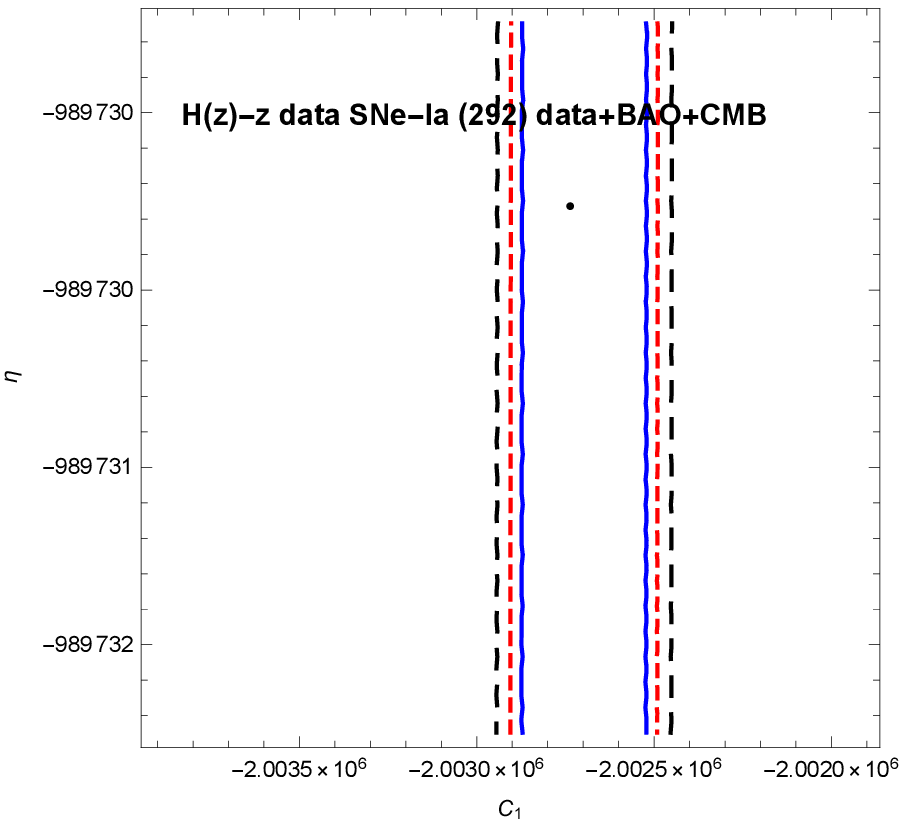}
\caption{Confidence contours for $\eta$ vs $C_1$ for the 292 SNe
Type-Ia +BAO +CMB data. }\label{f3.2} \vspace{4mm}
\end{minipage}~~\\
\vspace{0.5mm}
\end{figure}

Nowadays it is well known that the recent cosmic acceleration is
attributed to the presence of dark energy in the universe. But
although theoretically the concept seems to be relatively sound,
but we are yet to get any observational evidence of the same. As a
result, various investigations are underway. Of the many probes
the most interesting one is via the angular scale of the first
acoustic peak. Taking into account the angular scale of the sound
horizon at the surface of last scattering, encoded in the Cosmic
Microwave Background (CMB) power spectrum, CMB shift parameter is
defined by several authors (Bond, Efstathiou \& Tegmark 1997;
Efstathiou \& Bond 1999; Nesseris \& Perivolaropoulos 2007).
Although the parameter is insensitive to perturbations yet it is
suitable to constrain model parameters. The first peak of the CMB
power spectrum is basically a shift parameter given by
\begin{equation}\label{Eqn1.5.176}
{\cal R}=\sqrt{\Omega_{m}} \int_{0}^{z_{2}} \frac{dz}{E(z)}
\end{equation}
where $z_{2}$ is the value of redshift corresponding to the last
scattering surface. From the WMAP 7-year data available in the
work of Komatsu et al. (2011) the value of the parameter has been
obtained as ${\cal R}=1.726\pm 0.018$ at the redshift $z=1091.3$.
Now the $\chi^{2}$ function for the CMB measurement can be written
as
\begin{equation}\label{Eqn1.5.177}
\chi^{2}_{CMB}=\frac{({\cal R}-1.726)^{2}}{(0.018)^{2}}
\end{equation}
Now considering the three cosmological tests together, we can
perform the joint data analysis for CC/SNeTypeIa+BAO+CMB. The
total $\chi^{2}$ function for this case may be defined by
\begin{equation}\label{Eqn1.5.178}
\chi^{2}_{TOTAL}=\chi^{2}_{CC/SNe-Ia}+\chi^{2}_{BAO}+\chi^{2}_{CMB}
\end{equation}
As like before, the best fit values of the parameters $\eta$ and
$C_{1}$ of our model for joint analysis of BAO and CMB with CC \&
with SNe Type-Ia observational data have been evaluated and the
values have been presented in a tabular form in Table \ref{T5}.
66\% (solid, blue), 90\% (dashed, red) and 99\% (dashed, black)
$\eta$ Vs $C_{1}$ contours are also generated for this scenario
and are depicted in Figs. \ref{f3.1} and \ref{f3.2}. As expected
we see that the allowed range for the parameters lie in the
negative region. From \ref{f3.2} it is evident that the predicted
range for $C_{1}$ is highly constrained and tight.

\begin{table}[h!]
\centering \caption{CC + BAO + CMB Data \& SNe Type-Ia + BAO + CMB
Data}\label{T5}
\begin{tabular}{|c|c|c|c|}
\hline
~~~~$Data Type$ ~~~~~&~~~~$\eta$ ~~~~~& ~~~~~~~$C_{1}$ ~~~~~~~~& ~~~$\chi^{2}_{TOTAL-min}$~~~~~~\\
    \hline
    CC +BAO+CMB Data & -408.965 & -135.083 & 1874.33 \\
    \hline
     SNe Type-Ia +BAO+CMB Data & -989730.26 & -2002736.53 & 355032516.55 \\
    \hline
\end{tabular}

\vspace{2mm} \textbf{Table\ref{T5}:} Shows the best fit values of
$\eta$, $C_1$ and the minimum values of $\chi^{2}$ for fixed value
of other parameters for CC + BAO + CMB Data and SNe Type-Ia +BAO +
CMB data. \vspace{2mm}
\end{table}

\begin{figure}
\centering
\includegraphics[width=0.48\textwidth]{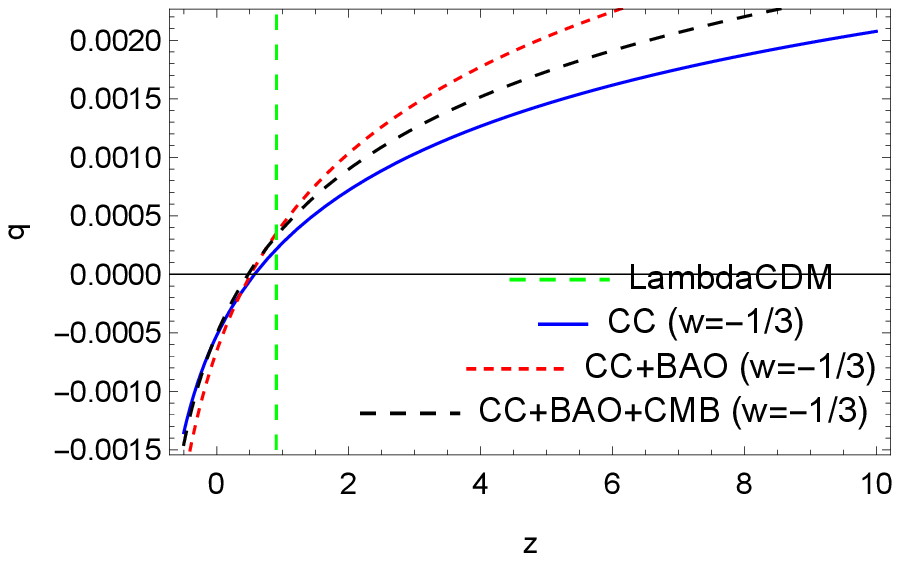}~~~~~\includegraphics[width=0.48\textwidth]{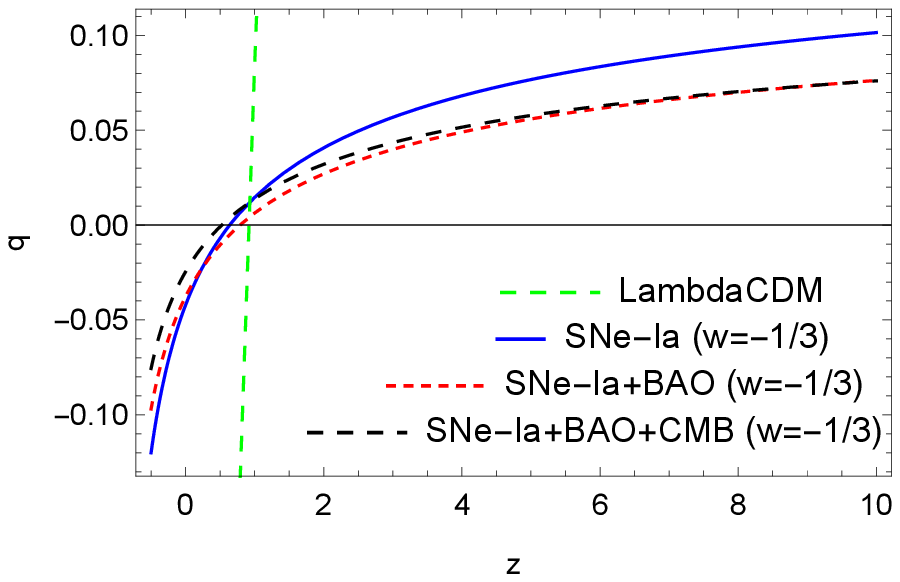}~~~~\\
Fig.7~~~~~~~~~~~~~~~~~~~~~~~~~~~~~~~~~~~~~~~~~~~~~~~~~~~~~~~Fig.8~~~~~~~~
\caption{The plot shows the deceleration parameter $q$ w.r.t. the
red-shift parameter $z$ using the constrained values of parameters
$\eta$ and $C_1$ from the CC data set.}\label{fig:figdec1}

\caption{The plot shows the deceleration parameter $q$ w.r.t. the
red-shift parameter $z$ using the constrained values of parameters
$\eta$ and $C_1$ from the SNe-Ia data set.} \label{fig:fig30}
\end{figure}

We obtained the plots for $q$ vs $z$ in figs.\ref{fig:figdec1} and
\ref{fig:fig30} using the constrained values of model parameters
for different data setting. We can see that the trajectories enter
the negative region around $z=0.6$ which shows the late time
accelerated expansion of the universe starting at $z\approx 0.6$.
Here the model fits perfectly with the observations. We have also
provided the plots for the $\Lambda$CDM case as reference. We see
that the plots for the $\Lambda$CDM case are vertical lines which
may be misleading on the first look. But actually they are not
vertical lines. In a larger scale they have variations with the
redshift just like the other curves. It is just because of the
scale that they look like vertical lines in the comparative
scenario with the plots for $w=-1/3$.

\subsection{Redshift-magnitude observations from Pantheon data Sample and Supernovae Type Ia: Union2 data sample}
\begin{figure}
\includegraphics[width=0.48\textwidth]{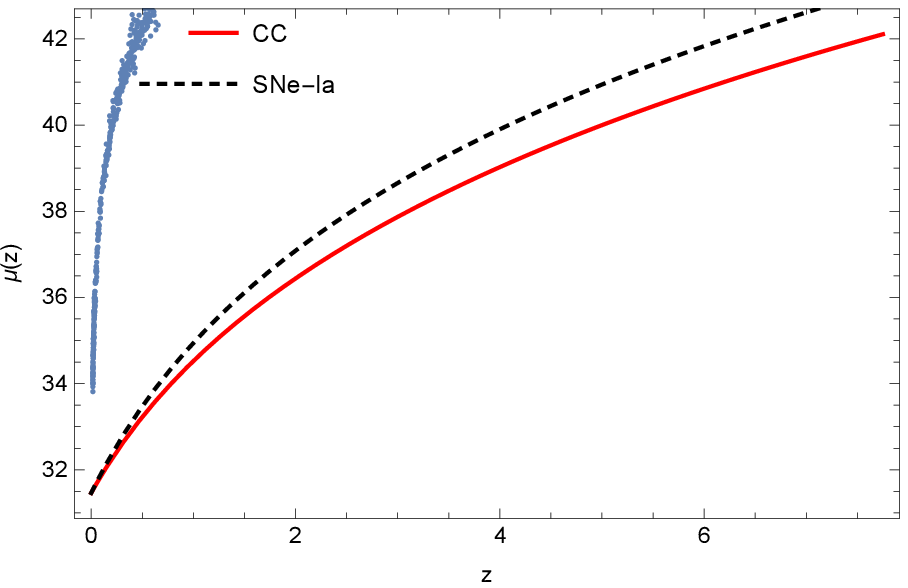}~~~~~\includegraphics[width=0.48\textwidth]{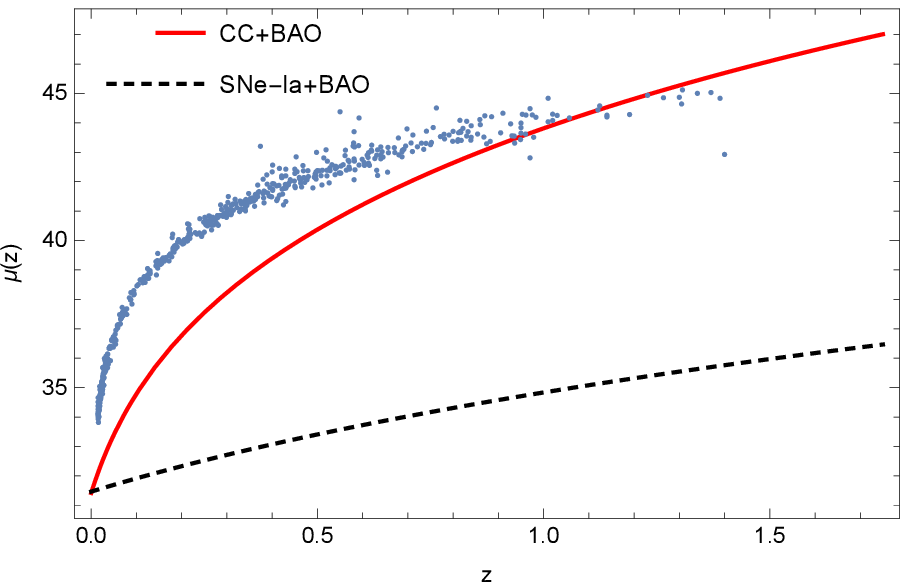}~~~~\\

~~~~~~~~~~~~~~~~~~~~~~~~Fig.9~~~~~~~~~~~~~~~~~~~~~~~~~~~~~~~~~~~~~~~~~~~~~~~~~~~~~~~~~~~~~~~Fig.10~~~~~~~\\
\caption{The plot shows the distance modulus $\mu(z)$ w.r.t. the
red-shift parameter $z$ using the constrained values of parameters
$\eta$ and $C_1$ from the CC and SNe-Ia data sets. The blue dots
show the Union2 sample points.}\label{fig:figsn0}

\caption{The plot shows the distance modulus $\mu(z)$ w.r.t. the
red-shift parameter $z$ using the constrained values of parameters
$\eta$ and $C_1$ from the CC+BAO and SNe-Ia+BAO data sets. The
blue dots show the Union2 sample points.} \label{fig:figsn1}
\end{figure}

\begin{figure}
\centering
\includegraphics[width=0.55\textwidth]{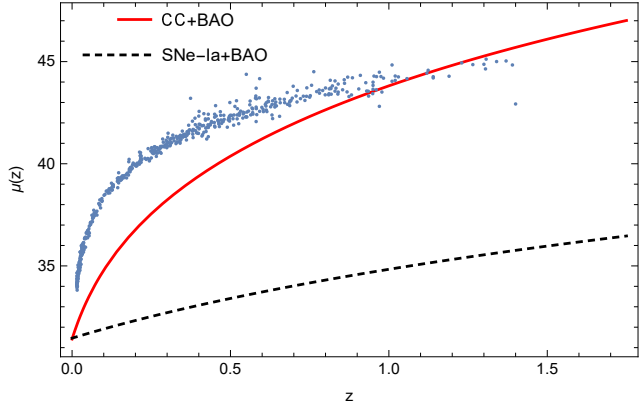}
\caption{The plot shows the distance modulus $\mu(z)$ w.r.t. the
red-shift parameter $z$ using the constrained values of parameters
$\eta$ and $C_1$ from the CC+BAO+CMB and SNe-Ia+BAO+CMB data sets.
The blue dots show the Union2 sample points.}
  \label{fig:figsn2}
\end{figure}

At the turn of the last century the Supernovae Type Ia
observations furnished satisfactory evidence for the fact that of
late the universe has actually entered a phase of accelerated
expansion. Since 1995, two teams of High redshift Supernova Search
and the Supernova Cosmology Project discovered several types of Ia
supernovas at the high redshifts (Perlmutter et al. 1998, 1999;
Riess et al. 1998, 2004). These observations triggered the concept
of dark energy as an explanation to the cosmic acceleration.
Further observations by Riess et al. (2007) and Kowalaski et al.
(2008) directly measured the distance modulus ($\mu(z)$) of
supernovae corresponding to high redshifts. A set of $557$ data
points including those from SNe-Ia known as Union2 sample is
reported by Amanullah et al. (2010). From these observations the
luminosity distance can be given by,
\begin{equation}
d_{L}(z)=(1+z)H_{0}\int_{0}^{z}\frac{dz'}{H(z')}
\end{equation}
where $z'$ is the independent variable in the above integral.
Moreover the distance modulus (which is defined as the difference
between the apparent magnitude and the absolute magnitude of any
astronomical object) is given by,
\begin{equation}
\mu(z)=5\log_{10}\left[\frac{d_{L}(z)/H_{0}}{1MPc}\right]+25
\end{equation}
In the figs. \ref{fig:figsn0}, \ref{fig:figsn1} and
\ref{fig:figsn2} we have obtained the best fit of the distance
modulus for our theoretical EMSG model for the constrained values
of parameters both from CC and SNe-Ia observations. These have
been compared with the union2 data sample represented by the blue
dots. In fig.\ref{fig:figsn0} we have used the pure CC and SNe-Ia
data to obtain the best fit. We can see that the best fits show
significant deviation from the union2 data sample, with CC data
showing a greater deviation. In \ref{fig:figsn1} the results
obtained from the joint analysis of CC+BAO and SNe-Ia+BAO have
been used to obtain the best fits. Here we see that the deviations
from the union2 data sample have considerably reduced and the fit
obtained from the CC+BAO is in agreement with the union2 sample
around $z=1$. But the fit from SNe-Ia+BAO data still has large
deviations from the union2 sample. In \ref{fig:figsn2} fits are
obtained from the joint analysis of CC+BAO+CMB and SNe-Ia+BAO+CMB.
Just like \ref{fig:figsn1} here also we witness a significant
agreement of CC+BAO+CMB fit with the union2 sample data. But the
fit from SNe-Ia+BAO+CMB is still significantly deviated from the
sample data. This analysis throws light on the two data sets in a
comparative scenario. In a comparative scenario with union2 sample
both CC data and SNe-Ia data show significant deviations, but when
joint analysis with BAO and CMB is considered, CC data shows
agreement with the union2 sample at around $z=1$.

\begin{figure}
\centering
\includegraphics[width=0.75\textwidth]{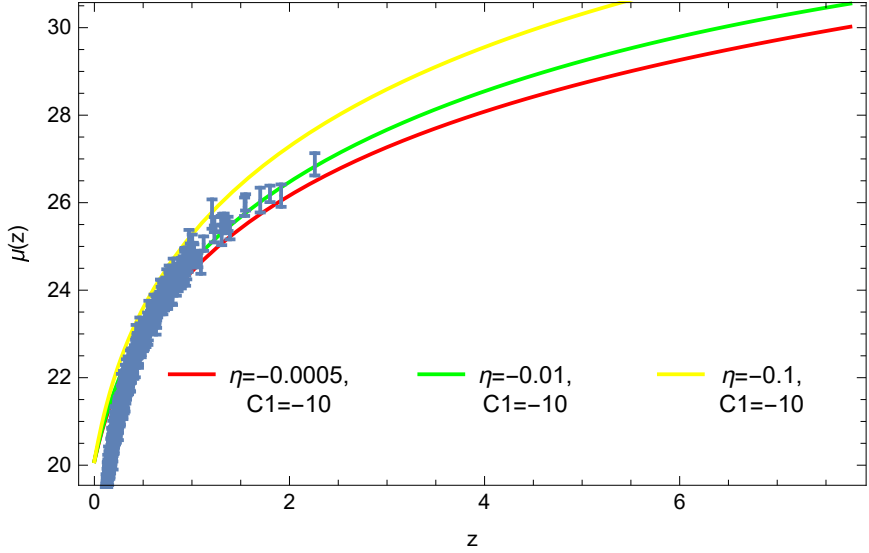}
\caption{The plot shows the distance modulus $\mu(z)$ w.r.t. the
red-shift parameter $z$ for different values of parameters $\eta$
keeping $C_1$ constant. The blue dots show the Pantheon sample
points.} \label{fig:figsn3}
\end{figure}

\begin{figure}
\centering
\includegraphics[width=0.75\textwidth]{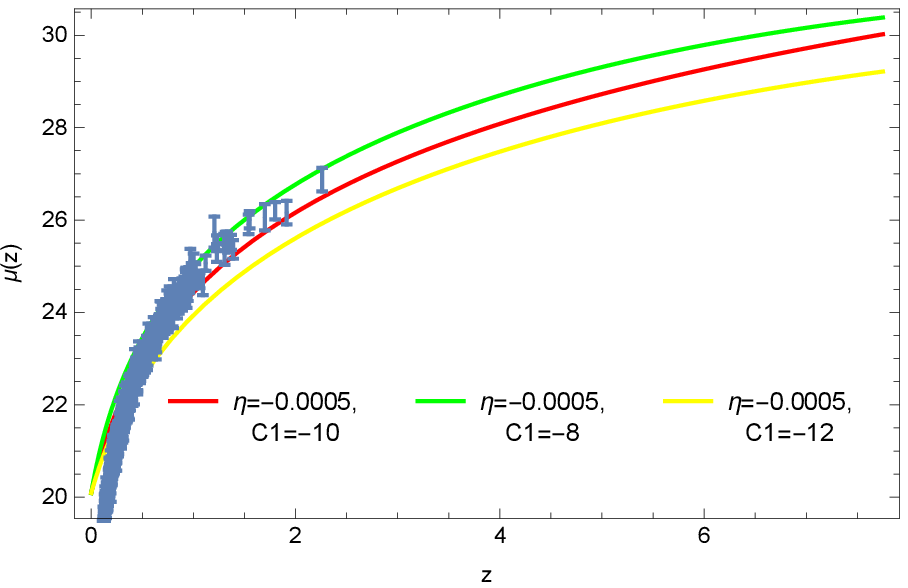}
\caption{The plot shows the distance modulus $\mu(z)$ w.r.t. the
red-shift parameter $z$ using different values of parameters $C_1$
keeping $\eta$ constant. The blue dots show the Pantheon sample
points.} \label{fig:figsn4}
\end{figure}

In figs.\ref{fig:figsn3} and \ref{fig:figsn4} we have plotted the
distance modulus of our model for different values of the
parameters $\eta$ and $C_{1}$, with the recently published
Pantheon data points in the background represented by the blue
cluster. In the fig.\ref{fig:figsn3} we have varied $\eta$ keeping
$C_{1}$ constant and in fig.\ref{fig:figsn4} we have varied
$C_{1}$ keeping $\eta$ constant. In both the figures the Pantheon
cluster with error bars actually helps us to get a comparative
outlook of our theory with the recent astronomical observations.
We can get a feel of the degree of compliance of our model with
the recent observations.

\section{Multi-component Universe model}
In the previous section we have considered a mono-component
universe model where the matter sector of the universe was
considered to be composed of a single component, given by a
cosmological fluid with EoS $w=-1/3$. But the observations favour
a multi-component matter universe. So in this section we will
consider dust in our model along with the cosmological fluid
content. It is well-known that the dust component of matter is
pressure-less and constitutes about $30\%$ of the universe. Most
of this portion is invisible in nature because it does not
interact with the electromagnetic field. Moreover this portion of
matter has minimum interaction with the baryonic component of the
universe and interacts only through gravitational interaction.
Here we consider the energy density of matter as
$\rho=\rho_{fluid}+\rho_{dust}$, where $\rho_{dust}$ and
$\rho_{fluid}$ represent the energy densities of dust and other
fluid component respectively. Moreover the total matter pressure
will remain the same $p=p_{fluid}$ because dust is pressureless,
i.e., $p_{dust}=0$. This will eventually make the equation state
of dust as $w_{dust}=p_{dust}/\rho_{dust}=0$. For this
two-component model we get $T_{\mu\nu (i)}T^{\mu\nu
(i)}=\rho_{i}^{2}\left(1+3w_{i}^2\right)$, where $i$ represents
the fluid component. So for dust we have $T_{\mu\nu (i)}T^{\mu\nu
(i)}=\rho_{dust}^{2}$ and for the other fluid component $T_{\mu\nu
(i)}T^{\mu\nu (i)}=\rho_{fluid}^{2}\left(1+3w_{fluid}^2\right)$.
The modified FLRW equations for this model take the forms,
\begin{eqnarray}
-3f_R\Big(\dot{H}+ H^2\Big)+\frac{f}{2}+3 H \dot{f_R}&=&\kappa^2\Big(\rho_{fluid}+\rho_{dust}+\frac{1}{\kappa^2}f_{\mathbf{T^2}}\Theta^2\Big)\,,\label{FW3} \\
-f_R(\dot{H}+3 H^2)+\frac{1}{2}f+\ddot{f_R}+2 H
\dot{f_R}&=&-\kappa^2 p_{fluid}\,,\label{FW4}
\end{eqnarray}
where
\begin{equation}
\mathbf{\Theta^2}:=\Theta_{\mu\nu}\Theta^{\mu\nu}=\left(\rho_{fluid}+\rho_{dust}\right)^2+4p
\left(\rho_{fluid}+\rho_{dust}\right)+3p_{fluid}^2\label{Theta2}
\end{equation}
Now the individual continuity equations for the different matter components are given below:\\

For cosmological fluid:\\

\begin{eqnarray}
\dot{\rho}_{fluid}+3H(w_{fluid}+1)\rho_{fluid}&=&-f_{\mathbf{T^2}} \left[3 \left(3 w_{fluid}^2+4 w_{fluid}+1\right) H \rho_{fluid} ^2+\left(3 w_{fluid}^2+8 w_{fluid}+1\right) \rho_{fluid} \dot{ \rho}_{fluid}\right]\nonumber\\
&&-\left(3 w_{fluid}^2+4 w_{fluid}+1\right) \rho_{fluid}^2
\dot{f}_{\mathbf{T^2}}\,.\label{consfluid}
\end{eqnarray}

For Dust:\\

\begin{equation}\label{consdust}
\dot{\rho}_{dust}+3H\rho_{dust}=-f_{\mathbf{T^2}}\left(3H\rho_{dust}^{2}+\rho_{dust}\dot{\rho}_{dust}\right)-\rho_{dust}^2
\dot{f}_{\mathbf{T^2}}
\end{equation}
Now the solution of the continuity equation for the fluid will be
similar to the one obtained in the previous section for the single
component fluid model. For the continuity equation of dust we see
that, it is difficult to get a solution in a model independent
way. So we proceed to get a solution for the model
$f(R,\mathbf{T^2})=R+\eta \mathbf{T^2}$. Solving the continuity
equation of dust for this model we get,
\begin{equation}\label{dustsol}
\rho_{dust}=C_{2}\left(1+z\right)^{3}
\end{equation}
where $C_{2}$ is the constant of integration. Now we can compare
the density components by studying their evolution with time. To
do this we consider the first FLRW equation as,
\begin{equation}\label{newflrw}
3H^{2}=\kappa^{2}\left(\rho_{dust}+\rho_{fluid}+\rho_{modified}\right)
\end{equation}
from the above equation we see that
$\Omega_{dust}+\Omega_{fluid}+\Omega_{modified}=1$, where
$\Omega_{dust}=\frac{\kappa^{2}\rho_{dust}}{3H^{2}}$,
$\Omega_{fluid}=\frac{\kappa^{2}\rho_{fluid}}{3H^{2}}$ and
$\Omega_{modified}=\frac{\kappa^{2}\rho_{modified}}{3H^{2}}$.
These are the dimensionless density components of the universe for
our model. We will study the evolution characteristics of these
density components.

\begin{figure}
\centering
\includegraphics[width=0.65\textwidth]{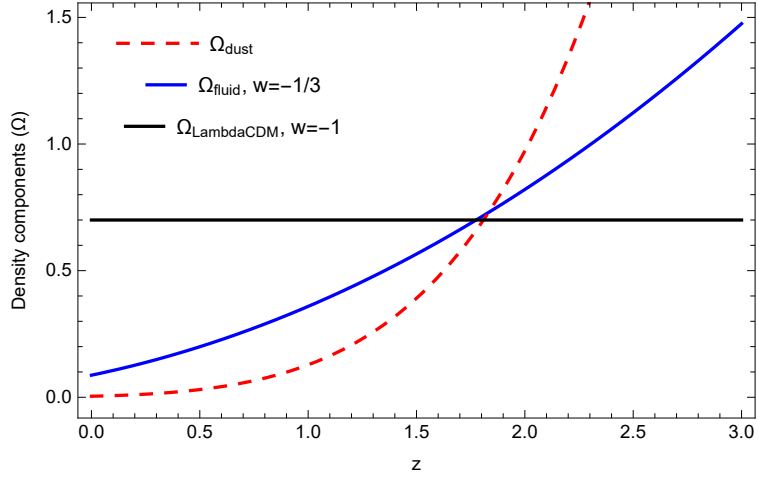}
\caption{The plot shows the evolution of density components
$\Omega$ w.r.t. the red-shift parameter $z$. Plot for the
$\Lambda$CDM case have been provided as reference. The initial
conditions are taken as $C_{1}=-135$, $\eta=-408$, $C_{2}=0.004$
and $\kappa=1$.} \label{fig:density1}
\end{figure}

\begin{figure}
\centering
\includegraphics[width=0.65\textwidth]{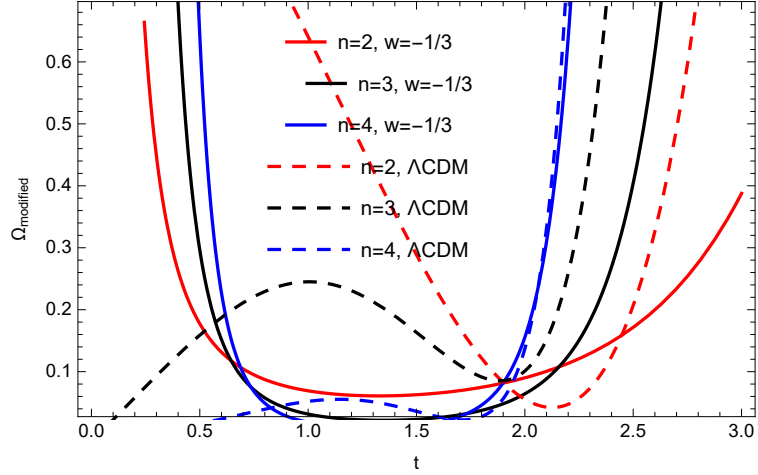}
\caption{The plot shows the evolution of the density component of
modified gravity $\Omega_{modified}$ w.r.t. time $t$ for different
values of parameters $n$. The corresponding scenarios for the
$\Lambda$CDM model have also been presented as reference. The
model parameters are taken as $C_{1}=-135$, $\eta=-408$,
$C_{2}=0.2$, $w=-1/3$, $z_{0}=0.1$ and $\kappa=1$.}
\label{fig:density2}
\end{figure}

\begin{figure}
\centering
\includegraphics[width=0.65\textwidth]{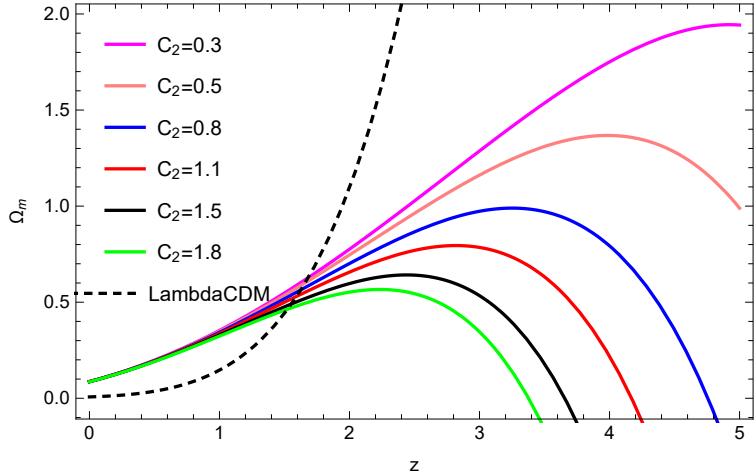}
\caption{The plot shows the evolution of the total matter density
component i.e. $\Omega_{m}=\Omega_{dust}+\Omega_{fluid}$  w.r.t.
redshift $z$ for different values of parameters $C_{2}$ for
$w=-1/3$. The case for $\Lambda$CDM have also been presented. The
model parameters are taken as $C_{1}=-135$, $\eta=-408$,
$w=-1/3$.} \label{fig:density3}
\end{figure}

In fig.\ref{fig:density1} we have plotted the evolution of the
density components of dust and matter fluid with $w=-1/3$ against
the redshift parameter $z$. Here we have considered the
constrained values of the parameters for the fluid. This helps us
to compare the evolution of a single component fluid model with
dust. Here the $C_{2}$ parameter is free and so by suitably fine
tuning the parameter we can get our model to follow the
observations and thus get an idea about correct the parametric
value. It should be stated here that both the matter components
are non-exotic in nature. In fig.\ref{fig:density2} we have
studied the behaviour of the density component arising from the
modification of gravity against time. Since our matter content
(dust and fluid) is non-exotic in nature, the effects coming from
the EMSG modifications to gravity should be responsible for the
accelerated expansion of the universe. Here we have used the
redshift-time dependence as $z(t)=z_{0}t^{n}$, where $z_{0}$ and
$n$ are constants. We have studied the evolution of
$\Omega_{modified}$ against time $t$ for different values of the
power $n$. It is seen that in the early times there is a
significant effect of the modification of gravity. The density
component due to EMSG modification peaks in the early universe and
wanes out with the evolution of time. This is in complete
correspondence with the work of Akarsu et. al 2018. Probably this
peak is responsible for the cosmological inflation in the early
universe. With the evolution of time $\Omega_{modified}$ decays
and universe is dominated by matter in the intermediate phase.
Finally again $\Omega_{modified}$ shoots up and dominates the
universe in late times, thus accounting for the recent cosmic
acceleration. It should be mentioned over here that the
contributions from the modified gravity are actually equivalent to
the exotic contributions from a dark energy model. So our
non-exotic matter along with the EMSG modifications gives quite a
reasonable proposition. In fact it is almost meaningless to
consider both exotic matter and contributions from modified
gravity because in such a scenario it is quite obvious that an
accelerated expansion will always be realized irrespective of
initial conditions. So our multi-component model is in complete
agreement with the evolution history of the universe. The overall
evolution of the system will be the combination of the
contributions from the three component: fluid, dust and modified
gravity.

\begin{figure}
\centering
\includegraphics[width=0.48\textwidth]{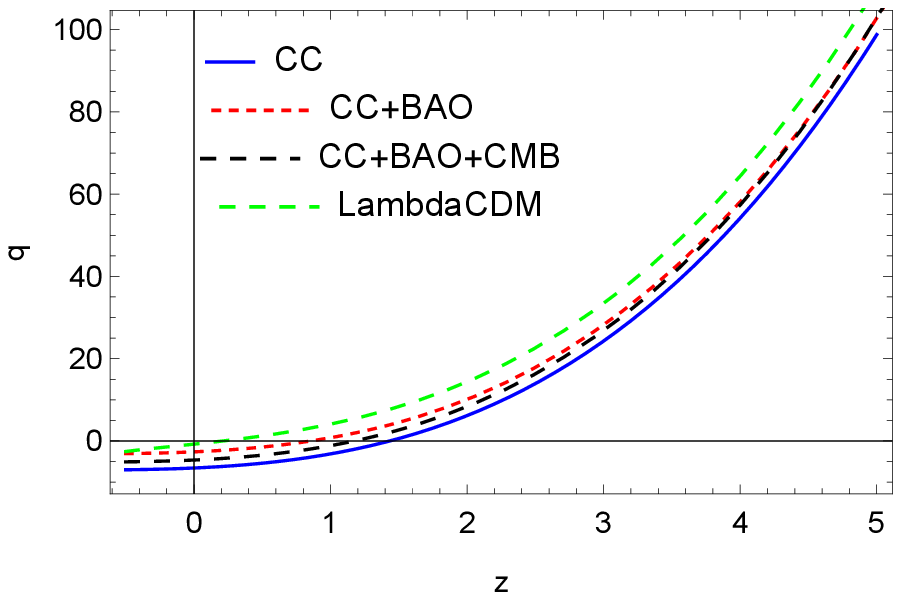}~~~~~\includegraphics[width=0.48\textwidth]{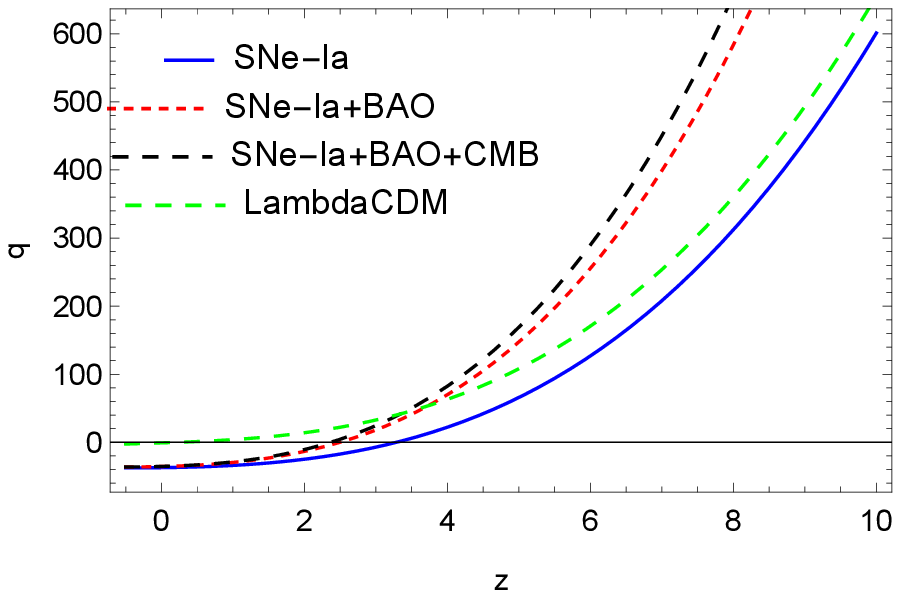}~~~~\\

~~~~~~~~~~~~~~~~Fig.17~~~~~~~~~~~~~~~~~~~~~~~~~~~~~~~~~~~~~~~~~~~~~~~~~~~~~~~Fig.18~~~~~~~~

\caption{The plot shows the deceleration parameter $q$ w.r.t. the
red-shift parameter $z$ using the constrained values of parameters
$\eta$ and $C_1$ from the CC data set for the multi-component
model.}\label{fig:figdec2}

\caption{The plot shows the deceleration parameter $q$ w.r.t. the
red-shift parameter $z$ using the constrained values of parameters
$\eta$ and $C_1$ from the SNe-Ia data set for the multi-component
model.} \label{fig:figdec3}
\end{figure}

In figure \ref{fig:density3} we have plotted the total matter
density components, i.e.,
$\Omega_{m}=\Omega_{dust}+\Omega_{fluid}$ against the redshift
$z$. Here the equation of state of the fluid is taken as $w=-1/3$
corresponding to the solution that we have got. So it is obvious
that both the components of $\Omega_{m}$ are relatively non-exotic
in nature. The acceleration is driven by the pure geometric
effects of modified gravity. From the observations we know that
the total share of $\Omega_{m}$ should be around 0.3. In our study
we see that the matter density decreases as the universe evolves,
which is obvious for an expanding universe. In
fig.\ref{fig:density3} we have obtained plots for different values
of the dust parameter $C_{2}$. We see that in almost all the cases
we can realize a scenario when $\Omega_{m} \approx 0.3$
corresponding to some value of redshift. From the figure we can
see that irrespective of the value of $C_{2}$ this is happening
for $z \approx 1$. Note that we have used the constrained values
of $\eta$ and $C_{1}$ that we have obtained from our observational
data analysis. Obviously by changing these parameters we can alter
the redshift value at which we realize $\Omega_{m} \approx 0.3$,
but since we have used the constrained values to obtain the plot,
it is reasonable to be satisfied with $z \approx 1$. This argument
actually helps us to address the degeneracy problem with the value
of $\Omega_{m}$ mentioned in the work Faria et. al 2019.  From the
observations of galaxy rotation curves and gravitational lensing
it is also seen that most of this matter is dark in nature. In all
the plots \ref{fig:density1}, \ref{fig:density2} and
\ref{fig:density3}, we have presented the scenario for the
$\Lambda$CDM model as a reference for the other trajectories. In
the figs.\ref{fig:figdec2} and \ref{fig:figdec3}, we have plotted
the deceleration parameter $q$ against the redshift parameter $z$
for our multi-component model. We see that for the both the
datasets we get late time accelerated expansion of the universe,
but it seems that the model fits better with the CC data. This is
because the origin of the accelerating phase is more pushed
towards the redshift value $z\approx 0.6$ for the CC data
(fig.\ref{fig:figdec2}) as compared to the SNe data
(fig.\ref{fig:figdec3}). From this, we can contemplate that the
constrained values of the model parameters of the EMSG model gives
more realistic results with the CC data. We have also presented
the trajectory for the $\Lambda$CDM case for comparison.

\section{Discussion and Conclusion}
In this work we have studied the increasingly popular $f(R,T^{2})$
theory. The theory has been constrained using observational data.
Two different data-sets have been used for this purpose, namely
the cosmic chronometer (CC) data and the Supernovae (SNe) Type-Ia
data (See Appendix for the full data table of CC data). It should
be mentioned that both these data are a set of $z-H(z)$ points,
which is the prime requirement of our data analysis mechanism. The
energy momentum squared gravity has a basic mathematical issue
regarding the integration of the continuity equation. The equation
cannot be integrated generically using the known mathematical
methods. In fact it was shown in Board \& Barrow (2017) that it
can be integrated for only two values of the barotropic parameter
viz. $w=-1$ and $w=-1/3$. Here we have integrated the continuity
equation for both these values and used the expression for energy
density obtained for $w=-1/3$ in the further computations. The
cosmological implications of our obtained solution is discussed in
detail and all possible limits of the solution is considered
including its reconciliation to GR as $\eta \rightarrow 0$.

The Hubble parameter $H$ is expressed in terms of the redshift
parameter $z$ in order to suit the nature of observational data
used. Then we have used this in the set-up to minimize the
$\chi^{2}$ distribution. In the $\chi^{2}$ variable we have
actually compared the theoretically obtained expression for the
Hubble parameter $H$ in terms of $z$ with the observed values of
$H$ corresponding to $z$ from the data. The best fit values of the
model parameters are then obtained from the $\chi^{2}$ function.
We see that the model has two free parameters $\eta$ and $C_{1}$.
Fitting the model with the data we have constrained these two
parameters and found out the most likely values of them. Further
we have used the peak parameters from BAO and CMB observations and
clubbed them with the two data-sets to further constrain the
parameters. For each data-set analysis have been done in three
steps. For the CC data, we have performed analysis with pure 30
points CC data, CC+BAO and CC+BAO+CMB. Similarly for SNe-Ia data
analysis have been performed with the 292 point SNe-Ia data,
SNe-Ia+BAO and SNe-Ia+BAO+CMB. The joint analysis is expected to
give far better results as far as the evolution of the universe is
concerned. For each set we have obtained the constrained values of
the free parameters $\eta$ and $C_{1}$. We have seen that both the
parameters acquire values in the negative range which is
completely consistent with a satisfactory cosmological behaviour.
We have also recorded the minimum value of the $\chi^{2}$
statistic, which is the basic statistical tool used in all the
analysis. For each analysis we have presented confidence contours
for the predicted values of the free parameters showing the
$66\%$, $90\%$ and $99\%$ confidence levels. From the confidence
contours we get clear idea about the values of the model
parameters along with their error bounds. The black dots in each
contour plots represent the best possible value of the parameters
as predicted by our statistical set-up correlating the
observational data and the theoretical model. Finally a best fit
of distance modulus is obtained for the theoretical EMSG model
both for CC and SNe-Ia data sets. It was seen that they show
significant deviations with the union2 sample data. However the
deviation for the CC+BAO and CC+BAO+CMB joint analysis is
considerably reduced and agreement with union2 sample is
established around $z=1$. Moreover we have also obtained a best
fit distance modulus of our model against the recently published
Pantheon data sample and tried to get an idea about the compliance
of the theoretical model with recent observations.

Finally we have studied a multi-component universe by adding dust
to our fluid model. Fluid with equation of state $w=-1/3$ is taken
with dust with equation of state $w=0$ to constitute the
non-exotic matter part of the universe. The acceleration in such a
model is supposed to be driven by the EMSG modifications to the
gravity characterized by the density component arising from the
modified gravity. We have studied the evolution of all the three
density components $\Omega_{dust}$, $\Omega_{fluid}$ and
$\Omega_{modified}$ in comparative scenarios. We have seen that
for suitable initial conditions our model gives the values which
are favoured by observations. We have also investigated the trends
of the deceleration parameter for the multi-component scenario.
All the results have been compared with those of the standard
$\Lambda$CDM model. To sum up, the present analysis discusses the
correlation of the theoretical EMSG model with observational data
and clearly is a significant development of the energy momentum
squared gravity model as well as our understanding of the
universe.\\\\\\

\section*{Acknowledgments}

Both C.R. and P.R. acknowledge the Inter University Centre for
Astronomy and Astrophysics (IUCAA), Pune, India for granting
Visiting Associateship. We thank the anonymous referee for his/her
valuable comments that helped us to improve the quality of the
manuscript considerably.\\\\

\section{Appendix}

See Table \ref{T1} for the $30$ point Cosmic Chronometer Data.
\begin{table}[h!]
\centering \caption{Cosmic Chronometer $30$ point Data
Set}\label{T1}
\begin{tabular}{|c|c|c|c|c|c|}
\hline
  ~~~~~~$z$ ~~~~& ~~~~$H(z)$ ~~~~~& ~~~~$\sigma(z)$~~~~~& ~~~~$z$ ~~~~& ~~~~$H(z)$ ~~~~~& ~~~~$\sigma(z)$~~~~\\
  \hline
  0.07 & 69 & $\pm$ 19.6 & 0.4783 & 80.9 & $\pm$ 9\\
  0.09 & 69 & $\pm$ 12 & 0.48 & 97 & $\pm$ 62\\
  0.12 & 68.6 & $\pm$ 26.2 &  0.593 & 104 & $\pm$ 13\\
  0.17 & 83 & $\pm$ 8 & 0.68 & 92 & $\pm$ 8\\
  0.179 & 75 & $\pm$ 4 & 0.781 & 105 & $\pm$ 12\\
  0.199 & 75 & $\pm$ 5 & 0.875 & 125 & $\pm$ 17\\
  0.2 & 72.9 & $\pm$ 29.6 & 0.88 & 90 & $\pm$ 40\\
  0.27 & 77 & $\pm$ 14 & 0.9 & 117 & $\pm$ 23\\
  0.28 & 88.8 & $\pm$ 36.6 &  1.037 & 154 & $\pm$ 20\\
  0.352 & 83 & $\pm$ 14 & 1.3 & 168 & $\pm$ 17\\
  0.3802 & 83 & $\pm$ 13.5 & 1.363 & 160 & $\pm$ 33.6\\
  0.4 & 95 & $\pm$ 17 & 1.43 & 177 & $\pm$ 18\\
  0.4004 & 77 & $\pm$ 10.2 &  1.53 & 140 & $\pm$ 14\\
  0.4247 & 87.1 & $\pm$ 11.2 & 1.75 & 202 & $\pm$ 40\\
  0.44497 & 92.8 & $\pm$ 12.9 &  1.965 & 186.5 & $\pm$ 50.4\\ \hline
\end{tabular}

\vspace{10mm} \text{Table~\ref{T1}: Shows the 30 point cosmic
chronometer $z-H(z)$ data with the standard error $\sigma(z)$.}
\end{table}


\begin{thebibliography}{99}

\bibitem{1} Akarsu O., Barrow J. D., Cikintoglu S., Eksi K. Y., Katirci N., 2018, Phys. Rev. D, 97, 12\\
Akarsu O., Katirci N., Kumar S., Nunes R. C., Sami M., 2018, Phys. Rev. D, 98, 6\\
Akarsu O, Katirci N., Kumar S., 2018, Phys. Rev. D, 97, 2\\
Akarsu O., Barrow, J. D., Board C. V. R., Uzun N. M., Vazquez J. A., 2019, Eur. Phys. J. C, 79, 846\\
Amanullah R. et al., 2010, Astrophys. J., 716, 712\\
Amendola L., Polarski D., Tsujikawa S., 2007, Phys. Rev. Lett., 98, 131302\\
Amendola L., Gannouji R,. Polarski D., Tsujikawa S., 2007, Phys. Rev. D, 75, 083504\\
Astier P. et al., 2006, Astron. Astrophys., 447, 31\\
Azizi T., Yaraie E., 2014, Int. J. Mod. Phys. D., 23, 1450021\\
Bahamonde S., Marciu M., Rudra P., 2019, Phys. Rev. D, 100, 8\\
Bertolami O., Boehmer C. G., Harko T., Lobo F. S. N., 2007, Phys. Rev. D, 75, 104016\\
Board C. V. R., Barrow J. D., 2017, Phys. Rev. D, 96, 12\\
Bond J. R., Efstathiou G., Tegmark M., 1997, Mon. Not. R. Astron. Soc., 291, L33\\
Brax P., 2018, Rep. Prog. Phys., 81, 1\\
Capozziello S., De Laurentis M., Faraoni V., 2010, Open Astron. J., 3, 49\\
Carroll S. M., Duvvuri V., Trodden M., Turner M. S., 2004, Phys. Rev. D, 70, 043528\\
Clifton T., Ferreira P. G., Padilla A., Skordis C., 2012, Phys. Rept., 513, 1\\
Cognola G., Elizalde E., Nojiri S., Odintsov S. D., Sebastiani L., Zerbini S., 2008, Phys. Rev. D, 77, 046009\\
De Felice A., Tsujikawa S., 2010, Living Rev. Relativ., 13, 3\\
Efstathiou G., Bond J. R., 1999, Mon. Not. R. Astron. Soc., 304, 75\\
Eisenstein D. J. et al., 2005, Astrophys. J., 633, 560\\
Faria M. C. F., Martins C. J. A. P, Chiti, F., Silva B. S. A., 2019, Astron. Astrophys., 625, A127\\
Farooq O., Ratra B., 2013, Astrophys. J., 766, L7\\
Ghose S., Thakur P., Paul B. C., 2012, Mon. Not. R. Astron. Soc., 421, 20\\
Gomez-Valent A., Amendola L., 2019, presentation at "15th Marcel Grossmann Meeting on Recent Developments in Theoretical and Experimental General Relativity, Astrophysics and relativistic Field Theories (MG15)" Rome, Italy, July 1-7, 2018\\
Haghani Z., Harko T., Lobo F. S. N., Sepangi H. R., Shahidi S., 2013, Phys. Rev. D, 88, 4\\
Harko T., 2008, Phys. Lett. B., 669, 376\\
Harko T., 2010, Phys. Rev. D, 81, 044021\\
Harko T., Lobo F. S. N., 2010, Eur. Phys. J. C, 70, 373\\
Harko T., Lobo F. S. N., Nojiri S., Odintsov S. D., 2011, Phys. Rev. D, 84, 024020\\
Jimenez R., Loeb A., 2002, Astrophys. J., 573, 37\\
Katirci N., Kavuk M., 2014, Eur. Phys. J. Plus, 129, 163\\
Keskin A., 2018, AIP Conf. Proc., 2042, 1\\
Komatsu E. et al., 2011, Astrophys. J. Suppl., 192, 18\\
Kowalaski M. et al., 2008, Astrophys. J., 686, 749\\
Martin J., 2012, C R Phys, 13, 566\\
Moraes P. H. R. S., Sahoo P. K., 2018, Phys. Rev. D, 97, 2\\
Moresco M., 2015, Mon. Not. R. Astron. Soc., 450, 1\\
Moresco M. et al., 2016, J. Cosmol. Astropart. Phys., 1605, 014\\
Nagpal R., Singh J. K., Beesham A., Shabani H, 2019, Ann. Phys., 405, 234\\
Nari N., Roshan M., 2018, Phys. Rev. D, 98, 2\\
Nojiri S., Odintsov S. D., 2006, Phys. Rev. D, 74, 086005\\
Nojiri S., Odintsov S. D., 2007, Int. J. Geom. Meth. Mod. Phys., 4, 115\\
Nojiri S., Odintsov S. D., Oikonomou V. K., 2017, Phys. Rept., 692, 1\\
Nesseris S., Perivolaropoulos L., 2007, J. Cosmol. Astropart. Phys., 0701, 018\\
Paliathanasis A., Tsamparlis M., Basilakos S., 2011, Phys. Rev. D, 84, 123514\\
Paliathanasis A., 2016, Class. Quantum Grav., 33, 7\\
Paul B. C., Thakur P., Ghose S., 2010, Mon. Not. R. Astron. Soc., 407, 415\\
Paul B. C., Ghose S., Thakur P., 2011, Mon. Not. R. Astron. Soc., 413, 686\\
Perlmutter S. et al. 1998, Nature, 391, 51\\
Perlmutter S. et al., 1999, Astrophys. J., 517, 565\\
Riess A. G. et al., 1998, Astron. J., 116, 1009\\
Riess A. J. et al., 2004, Astrophys. J., 607, 665\\
Riess A. J. et al., 2007, Astrophys. J., 659, 98\\
Roshan M., Shojai F., 2016, Phys. Rev. D, 94, 4\\
Simon J., Verde L., Jimenez R., 2005, Phys. Rev. D, 71, 123001\\
Song Y-S., Hu W., Sawicki I., 2007, Phys. Rev. D, 75, 044004\\
Sotiriou T. P., Faraoni V., 2010, Rev. Mod. Phys., 82, 451\\
Spergel D. N. et al., 2003, Astrophys. J. Suppl., 148, 175\\
Stern D., Jimenez R., Verde L., Kamionkowski M., Stanford S., 2010, J. Cosmol. Astropart. Phys., 02, 008\\
Thakur P., Ghose S., Paul B. C., 2009, Mon. Not. R. Astron. Soc., 397, 1935\\
Wu P., Yu H. W., 2007, Phys. Lett. B, 644, 16\\
Zhang C., Zhang H., Yuan S., Zhang T.-J., Sun Y.-C., 2014, Res. Astron. Astrophys., 14, 10\\
Zlatev I, Wang L.-M. Steinhardt P. J, 1999, Phys. Rev. Lett., 82, 896\\




\end{thebibliography}
\end{document}